  \documentclass[
  aps,
  prc,
  twocolumn,
  showpacs,
  superscriptaddress
  ]{revtex4-1}

  \usepackage[english]{babel}
  \usepackage[dvips]{color,graphicx}

  \usepackage{etoolbox}
  \makeatletter
  \patchcmd{\Ginclude@eps}{"#1"}{#1}{}{}
  \makeatother

  \usepackage{amsmath}    
  \usepackage{amssymb}
  \usepackage{multirow}
  \usepackage{tabularx}
  \usepackage{graphicx}   
  \usepackage{hyperref}   
  \usepackage{url}

  \def\F#1#2#3{$\dfrac{#1}{#2}\approx{#3}$}
  \def\chiNDF{$\dfrac{\chi^2}{\text{NDF}}$}

\allowdisplaybreaks 

\begin{document}

\title{Mean charged multiplicities in charged-current neutrino scattering \\
       on hydrogen and deuterium}
         
\author{Konstantin S. Kuzmin} 
\email{KKuzmin@theor.jinr.ru}
\affiliation{Bogoliubov Laboratory of Theoretical Physics,
             Joint Institute for Nuclear Research, RU-141980 Dubna, Russia}
\affiliation{Institute for Theoretical and Experimental Physics, 
             RU-117259 Moscow, Russia}
\author{Vadim A. Naumov} 
\email{VNaumov@theor.jinr.ru}
\homepage{http://theor.jinr.ru/~vnaumov/}
\affiliation{Bogoliubov Laboratory of Theoretical Physics,
             Joint Institute for Nuclear Research, RU-141980 Dubna, Russia}

\date{\today} 

\begin{abstract}
Available experimental data are analyzed to derive simple parametrizations
for the mean charged-hadron multiplicities in charged-current neutrino and
antineutrino interactions with hydrogen and deuterium targets. 
The obtained results can be used in the (anti)neutrino-induced hadronic
shower modeling. 
\end{abstract}



\maketitle

\section{Introduction}
\label{Sec:Introduction}

The average charged-secondary-hadron multiplicity in full phase space,
$\langle{n_{\text{ch}}}\rangle$, is one of the basic observables describing
the final-state evolution with energy and it is therefore an essential input
in the (anti)neutrino-induced hadronic shower modeling.
For example, the so-called ``AGKY model'' \cite{Yang:09}, the default
hadronization model in the Monte Carlo neutrino event generators NEUGEN\,3 \cite{NEUGEN}
and GENIE-2.0.0 \cite{Andreopoulos:2006-2009,Andreopoulos:2009rq}, uses as starting point,
the well-known empirical expression
\begin{equation}
\label{StandardParametrization}
\langle{n_{\text{ch}}}\rangle=a+b\ln{W^2},
\end{equation}
in which $W$ is the invariant mass of the final-state hadrons (including neutrals) and the
coefficients $a$ and $b$ dependent on the initial state ($\nu/\overline{\nu}$ and struck nucleon) 
are determined by selected hydrogen and deuterium bubble-chamber experiments
and are treated as tuning parameters.
There are, however, serious reasons which suggest to refine the parametrization of
$\langle{n_{\text{ch}}}\rangle$ currently used in the neutrino generators.

The first reason is that the available data on charged multiplicity obtained
in different bubble-chamber experiments are generally in rather poor agreement with each
other (see, e.g., review papers \cite{Albini:75,Schmitz:80,Schmitz:81a,Matteuzzi:81,Kiselev:88,Schmitz:88-93}
and Table~\ref{Tab:MCHMCC_collection} in the next section).
It is therefore necessary to sort out the existing data in order to select the mutually
consistent and robust data sets, acceptable for statistical analysis.
The second trivial reason is that the simple linear parametrization \eqref{StandardParametrization}
does not work in the low-multiplicity region for the reactions with nonzero total hadronic charge $Q_h$,
since charged multiplicities lower than $|Q_h|$ cannot occur, while the expression~\eqref{StandardParametrization}
with the experimentally fitted parameters $a$ and $b$ cannot be extrapolated to the near-threshold values
of $W$ without violating this rule or even the positivity constraint.

Besides that, an accurate parametrization of $\langle{n_{\text{ch}}}\rangle$ in the low- and mid-$W$
regions is of central practical importance for the correct determination of the ``boundary'' in the
phase space between the exclusive (resonance) and inclusive (deep inelastic) contributions into the total
${\nu}N$ and $\overline{\nu}_{\mu}N$ cross sections \cite{Wcut,Kuzmin:2006dt} or into the simulated
count rates in the experiments with accelerator and atmospheric (anti)neutrino beams \cite{T2K-SK}. 

The strictly practical aim of this paper is to provide an economical parametrization of the charged-hadron
multiplicities for the charged-current induced $\nu_{\mu}p$, $\nu_{\mu}n$, $\overline{\nu}_{\mu}p$ and
$\overline{\nu}_{\mu}n$ reactions, valid in the whole kinematic region of $W$,
by using the available consistent data from the experiments performed on hydrogen and deuterium targets.
We do not discuss here the experiments on heavy nuclear targets, as well as more detailed data, such as
forward/backward or positive/negative hadron multiplicity asymmetries and so on.

\section{The data selection }
\label{Sec:Experimental_data}

The mean charged-hadron multiplicity in the muon neutrino and antineutrino
charged-current reactions on hydrogen and deuterium has been measured
in the Fermilab experiments 
E31 \cite{Derrick:76,Singer:77,Derrick:78,Barish:78,Derrick:80,Derrick:81,Derrick:82},
E45 \cite{Coffin:75,Chapman:76,Bell:79,VanderVelde:79} 
and 
E545 \cite{Kitagaki:1980pn,Kitagaki:80,Zieminska:83} 
with the 15-foot Bubble Chamber and in the CERN experiments
WA21 \cite{Saarikko:79,Schmitz:79,Allen:81,Schmitz:81b,Grassler:83,Jones:90,Jones:92}
and
WA25 \cite{Allasia:80,Barlag:81,Barlag:82,Allasia:84,Jongejans:89}
with the Big European Bubble Chamber (BEBC).
The data obtained with the FNAL and BNL hydrogen bubble chambers before 1976
are gathered in Ref.~\cite{Albini:75}. 

Table~\ref{Tab:MCHMCC_collection} summarizes the results of Refs.
\cite{Derrick:76,Singer:77,Derrick:78,Derrick:82,Coffin:75,Chapman:76,Bell:79,Kitagaki:80,Zieminska:83,%
      Saarikko:79,Schmitz:79,Allen:81,Grassler:83,Jones:90,Jones:92,Allasia:80,Barlag:81,Barlag:82},
represented in terms of the intercept and slope coefficients $a$ and $b$ of Eq.~\eqref{StandardParametrization}. 
We selected mainly the experiments in which no additional kinematic cuts were applied for determination
of the coefficients, but for comparison we also show several fits obtained under particular conditions,
which are indicated in the fifth column of Table~\ref{Tab:MCHMCC_collection}, where $Q^2$ is the 4-momentum
transfer squared and $y=1-E_{\mu}/E_{\nu,\overline{\nu}}$ is the usual scaling variable.
\begin{table*}[htb]
\caption{Values of the intercept $a$ and slope $b$ obtained in different experiments on
         charged-current ${\nu}_{\mu}$ and $\overline{\nu}_{\mu}$ scatterings on hydrogen
         and deuterium targets, by fitting the mean charged-hadron multiplicity
         $\langle{n_{\text{ch}}\rangle}$ to the relation \protect\eqref{StandardParametrization}
         within the $W^2$ ranges shown in fourth column ($W$ in GeV).
         Specific cut conditions applied in some experiments are shown in fifth column.
}
\label{Tab:MCHMCC_collection} 
\center{
\begin{tabularx}{\linewidth}{l>{\centering\arraybackslash}c>{\centering\arraybackslash}X>{\centering\arraybackslash}X>{\centering\arraybackslash}c>{\centering\arraybackslash}X>{\centering\arraybackslash}X}
 \hline\hline\noalign{\smallskip}
 Author(s), experiment, publ.\ date               & Ref.                & Target & $W^2$ range & Kinematic cuts     & Intercept $a$       & Slope $b$                                       \cr
 \noalign{\smallskip}
 \hline\noalign{\smallskip} 
 \multicolumn{7}{c}{$\nu_{\mu}p\to\mu^{-}X^{++}$}                                                                                                                                           \cr
 \noalign{\smallskip}
 \hline\noalign{\smallskip}
 Coffin \emph{et al.},       FNAL E45, 1975       & \cite{Coffin:75}    & H      & $4-200$     &                    & $ 1.0   \pm 0.3  $  & $1.1   \pm 0.1  $                               \cr
 \noalign{\smallskip}
 Chapman \emph{et al.},      FNAL E45, 1976       & \cite{Chapman:76}   & H      & $4-200$     &                    & $ 1.09  \pm 0.38 $  & $1.09  \pm 0.03 $                               \cr
 \noalign{\smallskip}
 Bell \emph{et al.},         FNAL E45, 1979       & \cite{Bell:79}      & H      & $4-100$     & $Q^2=2-64$~GeV$^2$ & $-               $  & $1.35  \pm 0.15 $                               \cr
 \noalign{\smallskip}
 Kitagaki \emph{et al.},     FNAL E545, 1980      & \cite{Kitagaki:80}  & $^2$H  & $1-100$     &                    & $ 0.80  \pm 0.10 $  & $1.25  \pm 0.04 $                               \cr
 \noalign{\smallskip}
 Zieminska \emph{et al.},    FNAL E545, 1983      & \cite{Zieminska:83} & $^2$H  & $4-225$     &                    & $ 0.50  \pm 0.08 $  & $1.42  \pm 0.03 $                               \cr
 \noalign{\smallskip}
 Saarikko \emph{et al.},     CERN WA21, 1979      & \cite{Saarikko:79}  & H      & $3-200$     &                    & $ 0.68  \pm 0.04 $  & $1.29  \pm 0.02 $                               \cr
 \noalign{\smallskip}
 Schmitz,                    CERN WA21, 1979      & \cite{Schmitz:79}   & H      & $4-140$     &                    & $ 0.38  \pm 0.07 $  & $1.38  \pm 0.03 $                               \cr
 \noalign{\smallskip}
 Allen \emph{et al.},        CERN WA21, 1981      & \cite{Allen:81}     & H      & $4-200$     &                    & $ 0.37  \pm 0.02 $  & $1.33  \pm 0.02 $                               \cr
 \noalign{\smallskip}
 Gr\"{a}ssler \emph{et al.}, CERN WA21, 1983      & \cite{Grassler:83}  & H      & $11-121$    &                    & $-0.05  \pm 0.11 $  & $1.43  \pm 0.04 $                               \cr
 \noalign{\smallskip}
 Jones \emph{et al.},        CERN WA21, 1990      & \cite{Jones:90}     & H      & $16-196$    &                    & $ 0.911 \pm 0.224$  & $1.131 \pm 0.086$                               \cr
 \noalign{\smallskip}
 Jones \emph{et al.},        CERN WA21, 1992      & \cite{Jones:92}     & H      & $9-200$     &                    & $ 0.40  \pm 0.13 $  & $1.25  \pm 0.04 $                               \cr
 \noalign{\smallskip}
 Allasia \emph{et al.},      CERN WA25, 1980      & \cite{Allasia:80}   & $^2$H  & $2-60$      &                    & $ 1.07  \pm 0.27 $  & $1.31  \pm 0.11 $                               \cr
 \noalign{\smallskip}
 Allasia \emph{et al.},      CERN WA25, 1984      & \cite{Allasia:84}   & $^2$H  & $8-144$     & $Q^2>1$~GeV$^2$    & $ 0.13  \pm 0.18 $  & $1.44  \pm 0.06 $                               \cr
 \noalign{\smallskip}
 \hline\noalign{\smallskip}
 \multicolumn{7}{c}{$\overline{\nu}_{\mu}p\to\mu^{+}X^{0}$}                                                                                                                                 \cr
 \noalign{\smallskip}
 \hline\noalign{\smallskip}
 Derrick \emph{et al.},      FNAL E31, 1976       & \cite{Derrick:76}   & H      & $4-100$     & $y>0.1$            & $ 0.04  \pm 0.37 $  & $1.27  \pm 0.17 $                               \cr
 \noalign{\smallskip}
 Singer,                     FNAL E31, 1977       & \cite{Singer:77}    & H      & $4-100$     & $y>0.1$            & $ 0.78  \pm 0.15 $  & $1.03  \pm 0.08 $                               \cr
 \noalign{\smallskip}
 Derrick \emph{et al.},      FNAL E31, 1978       & \cite{Derrick:78}   & H      & $1-50$      &                    & $ 0.06  \pm 0.06 $  & $1.22  \pm 0.03 $                               \cr
 \noalign{\smallskip}
 Derrick \emph{et al.},      FNAL E31, 1982       & \cite{Derrick:82}   & H      & $4-100$     & $0.1<y<0.8$        & $-0.44  \pm 0.13 $  & $1.48  \pm 0.06 $                               \cr
 \noalign{\smallskip}
 Gr\"{a}ssler \emph{et al.}, CERN WA21, 1983      & \cite{Grassler:83}  & H      & $11-121$    &                    & $-0.56  \pm 0.25 $  & $1.42  \pm 0.08 $                               \cr
 \noalign{\smallskip}
 Jones \emph{et al.},        CERN WA21, 1990      & \cite{Jones:90}     & H      & $16-144$    &                    & $ 0.222 \pm 0.362$  & $1.117 \pm 0.141$                               \cr
 \noalign{\smallskip}
 Jones \emph{et al.},        CERN WA21, 1992      & \cite{Jones:92}     & H      & $9-200$     &                    & $-0.44  \pm 0.20 $  & $1.30  \pm 0.06 $                               \cr
 \noalign{\smallskip}
 Allasia \emph{et al.},      CERN WA25, 1980      & \cite{Allasia:80}   & $^2$H  & $7-50$      &                    & $ 0.55  \pm 0.29 $  & $1.15  \pm 0.10 $                               \cr
 \noalign{\smallskip}
 Barlag \emph{et al.},       CERN WA25, 1981      & \cite{Barlag:81}    & $^2$H  & $6-140$     &                    & $ 0.18  \pm 0.20 $  & $1.23  \pm 0.07 $                               \cr
 \noalign{\smallskip}
 Barlag \emph{et al.},       CERN WA25, 1982      & \cite{Barlag:82}    & $^2$H  & $6-140$     &                    & $ 0.02  \pm 0.20 $  & $1.28  \pm 0.08 $                               \cr
 \noalign{\smallskip}
 Allasia \emph{et al.},      CERN WA25, 1984      & \cite{Allasia:84}   & $^2$H  & $8-144$     & $Q^2>1$~GeV$^2$    & $-0.29  \pm 0.16 $  & $1.37  \pm 0.06 $                               \cr
 \noalign{\smallskip}
 \hline\noalign{\smallskip}
 \multicolumn{7}{c}{$\nu_{\mu}n\to\mu^{-}X^{+}$}                                                                                                                                            \cr
 \noalign{\smallskip}
 \hline\noalign{\smallskip}
 Kitagaki \emph{et al.},     FNAL E545, 1980      & \cite{Kitagaki:80}  & $^2$H  & $1-100$     &                    & $ 0.21  \pm 0.10 $  & $1.21  \pm 0.04 $                               \cr
 \noalign{\smallskip}
 Zieminska \emph{et al.},    FNAL E545, 1983      & \cite{Zieminska:83} & $^2$H  & $4-225$     &                    & $-0.20  \pm 0.07 $  & $1.42  \pm 0.03 $                               \cr
 \noalign{\smallskip}
 Allasia \emph{et al.},      CERN WA25, 1980      & \cite{Allasia:80}   & $^2$H  & $2-60$      &                    & $ 0.28  \pm 0.16 $  & $1.29  \pm 0.07 $                               \cr
 \noalign{\smallskip}
 Allasia \emph{et al.},      CERN WA25, 1984      & \cite{Allasia:84}   & $^2$H  & $8-144$     & $Q^2>1$~GeV$^2$    & $ 1.75  \pm 0.12 $  & $1.31  \pm 0.04 $                               \cr
 \noalign{\smallskip}
\hline\noalign{\smallskip}
 \multicolumn{7}{c}{$\overline{\nu}_{\mu}n\to\mu^{+}X^{-}$}                                                                                                                                 \cr
 \noalign{\smallskip}
 \hline\noalign{\smallskip}
 Allasia \emph{et al.},      CERN WA25, 1980      & \cite{Allasia:80}   & $^2$H  & $7-50$      &                    & $ 0.10  \pm 0.28 $  & $1.16 \pm 0.10 $                                \cr
 \noalign{\smallskip} 
 Barlag \emph{et al.},       CERN WA25, 1981      & \cite{Barlag:81}    & $^2$H  & $4-140$     &                    & $ 0.79  \pm 0.09 $  & $0.93 \pm 0.04 $                                \cr
 \noalign{\smallskip}
 Barlag \emph{et al.},       CERN WA25, 1982      & \cite{Barlag:82}    & $^2$H  & $2-140$     &                    & $ 0.80  \pm 0.09 $  & $0.95 \pm 0.04 $                                \cr
 \noalign{\smallskip}
 Allasia \emph{et al.},      CERN WA25, 1984      & \cite{Allasia:84}   & $^2$H  & $8-144$     & $Q^2>1$~GeV$^2$    & $ 0.22  \pm 0.21 $  & $1.08 \pm 0.06 $                                \cr
 \noalign{\smallskip}
 \hline\hline\noalign{\smallskip}
\end{tabularx}}
\end{table*}
The intermediate data from Refs.~\cite{Coffin:75,Saarikko:79,Schmitz:79,Kitagaki:80,Barlag:81}
are included for completeness only.

The fits shown in Table \ref{Tab:MCHMCC_collection} were performed for different intervals of $W$
spanning the region from 1 to about 15~GeV, but typically lying above the resonance region ($W\lesssim2$~GeV). 
The quoted errors for $a$ and $b$ are statistical only, except for the result of Ref.~\cite{Jones:90}, where
the statistical and systematic errors are added in quadrature. A misprint in the value of $a$, reported
in Ref.~\cite{Zieminska:83} for the reaction $\nu_{\mu}p\to\mu^{-}X^{++}$ is corrected according to
Ref.~\cite{Nowak:2006xv}.
The values quoted from Ref.~\cite{Allasia:84} are recalculated from the $a$ and $b$ values obtained by fitting to
the charged multiplicities in the forward and backward hemispheres separately.
The intercept value is not reported in Ref.~\cite{Bell:79};
our estimation yields $a=0.30\pm0.51$ and $b=1.36\pm0.17$ with $\chi^2/\text{NDF}=0.3$.

As is seen from Table~\ref{Tab:MCHMCC_collection}, the results of individual
experiments and even of different sets of runs or data subsets and $W$ ranges used
in the successive analyses of the same experimental data and for the same reaction
vary by amounts greatly in excess of the quoted errors.
This is especially true for the intercept coefficient which varies, sometimes even changing sign,
within the wide ranges specified by the reaction. The discrepancies cannot be fully attributed
to the targets employed in the experiments and they insignificantly correlate with
the $W$ ranges of fittings, mean (anti)neutrino beam energies (not shown in the table),
and even with the used kinematic cutoffs. 

In an effort to extract more certain information on $\langle{n_{\text{ch}}\rangle}$
from the available data, it is instructive to use the summary statistics of the consistent
independent measurements. For further analysis, we selected the statistically reliable
experiments whose results were not revised in subsequent years. Namely we include into
the statistical analysis the data from Refs.~\cite{Derrick:78,Chapman:76,Zieminska:83,Jones:92,Jongejans:89}
and acceptable subsamples of data from Refs.~\cite{Jones:90,Allen:81,Barlag:82}.
There are several comments that we would like to make regarding the reasons for this choice.

First, we do not use intermediate or apparently obsolete results
(e.g., Refs.~\cite{Albini:75,Coffin:75,Saarikko:79,Schmitz:79,Kitagaki:80,Barlag:81})
and the reports presented only the resulting fits of $\langle{n_{\text{ch}}\rangle}$ rather
than the ``raw'' data points. 
Also, we cannot utilize the data obtained after imposing the stringent kinematic cuts
(e.g., Refs.~\cite{Derrick:76,Singer:77,Derrick:80,Derrick:81,Derrick:82,Kitagaki:1980pn,Allasia:84})
systematically distorting the value of $\langle{n_{\text{ch}}\rangle}$. 
A representative example is provided by the FNAL E31 experiment ($\overline{\nu}_{\mu}p$).
The data of the E31 experiment for the full phase space \cite{Derrick:78} are based on about
20\% of the final data sample \cite{Derrick:82}, but in the latter analysis, only such events
were selected which satisfy rather hard constraint $0.1<y<0.8$.
Hence we are forced to use the lower statistics data from Ref.~\cite{Derrick:78} obtained
within the lower $W$ range ($1<W^2<50$~GeV$^2$).

\begin{figure}[htb]
 \includegraphics[width=0.965\linewidth]{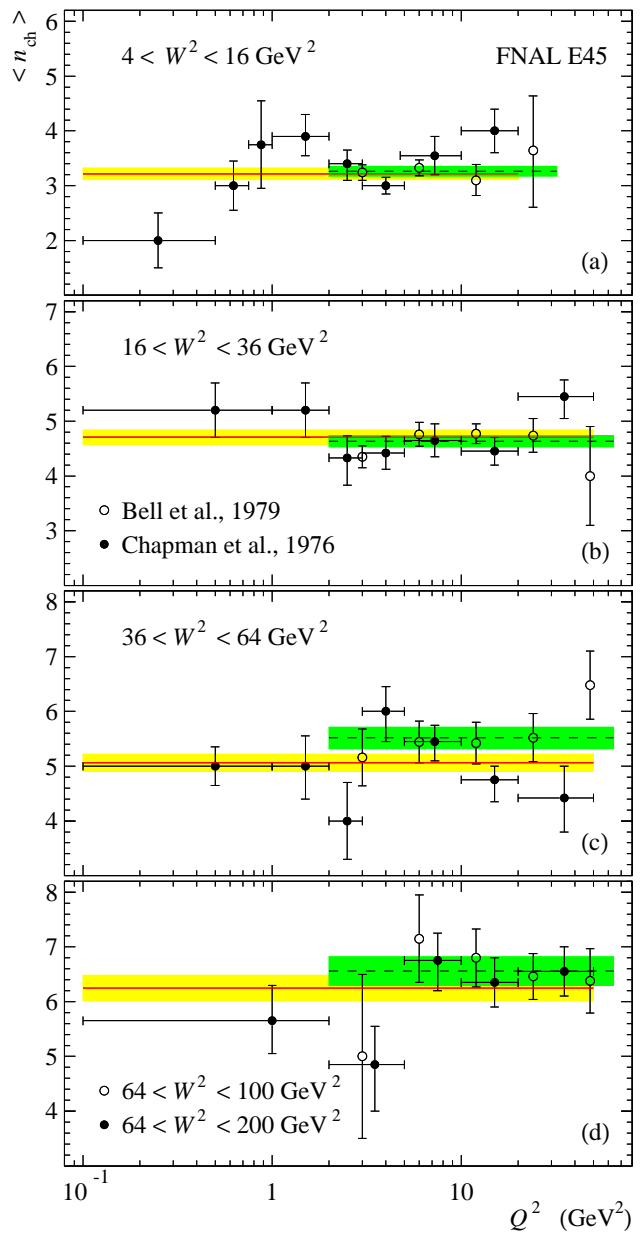}
 \caption{(Color online)
          Mean multiplicities of charged hadrons as a function of $Q^2$
          for various $W$ slices measured by the neutrino-hydrogen experiment E45 at FNAL.
          The filled circles are from Ref.~\cite{Chapman:76} and the open circles
          are from Ref.~\cite{Bell:79}. The errors shown are statistical only.
          The two lowest $W$ slices from Ref.~\cite{Bell:79} are merged into the single one (a). 
          The solid and dashed lines represent the overall averages for each slice using
          the data points from Refs.~\protect\cite{Chapman:76} and \protect\cite{Bell:79}, respectively.
          The filled bands display the 68\% confidence intervals around the estimated averages.
         }
 \label{Fig:MCHMCC_Q2_mn_p_FNAL_E45}
 \end{figure}

The final result of the neutrino-hydrogen experiment FNAL E45 \cite{Bell:79} (see also Ref.~\cite{VanderVelde:79})
is based on a data sample reduced by the conditions $2<W<10$~GeV and $2<Q^2<64$~GeV$^2$ and
presented as a dependence of $\langle{n_{\text{ch}}\rangle}$ on $Q^2$ for five narrow $W$ bins
located above the region of low-lying baryon resonances in exclusive channels.
Although this result partially supersedes the earlier data subset of FNAL E45 \cite{Chapman:76}
at $W>2$~GeV, the kinematic cuts used in Refs.~\cite{Bell:79} are at variance to our purposes.
So we have to use the low-statistics data set from Ref.~\cite{Chapman:76} not distorted by the cuts
and given as a $W$-dependency of $\langle{n_{\text{ch}}\rangle}$ for $1 \lesssim W \lesssim 14$~GeV.
A comparison of the results from Refs.~\cite{Chapman:76} and \cite{Bell:79} is plotted
in Fig.~\ref{Fig:MCHMCC_Q2_mn_p_FNAL_E45} ($\langle{n_{\text{ch}}\rangle}$ vs.\ $Q^2$ for several $W$ slices).
For this comparison we combined the two lowest $W$ bins $2-3$ and $3-4$~GeV used in
Ref.~\cite{Bell:79} into one bin $2-4$~GeV, which is shown in panel (a) of the figure.
The figure also shows the averages evaluated by fitting the data in each bin to a constant.
It is seen that, despite the different data sampling, $Q^2$ ranges of averaging and oscillations
of the data points, the averages for each bin derived from the two samples have similar uncertainties
and are in agreement within one or two standard deviations, thus testifying that the data
of Ref.~\cite{Chapman:76} remain appropriate for use.
Let us note at once that using these data only moderately affect our final results owing to
relatively large errors in comparison with the subsequent experiments with higher statistics. 

The analysis of Ref.~\cite{Jones:90} for both ${\nu}p$ and $\overline{\nu}p$ reactions
is based on the same data sample as in Ref.~\cite{Jones:92} and the main part of the results
of Ref.~\cite{Jones:90} relevant to our study is presented in Ref.~\cite{Jones:92}, except
for the data at the lowest invariant hadronic masses.

\begin{figure}[hbt]
 \includegraphics[width=0.98\linewidth]{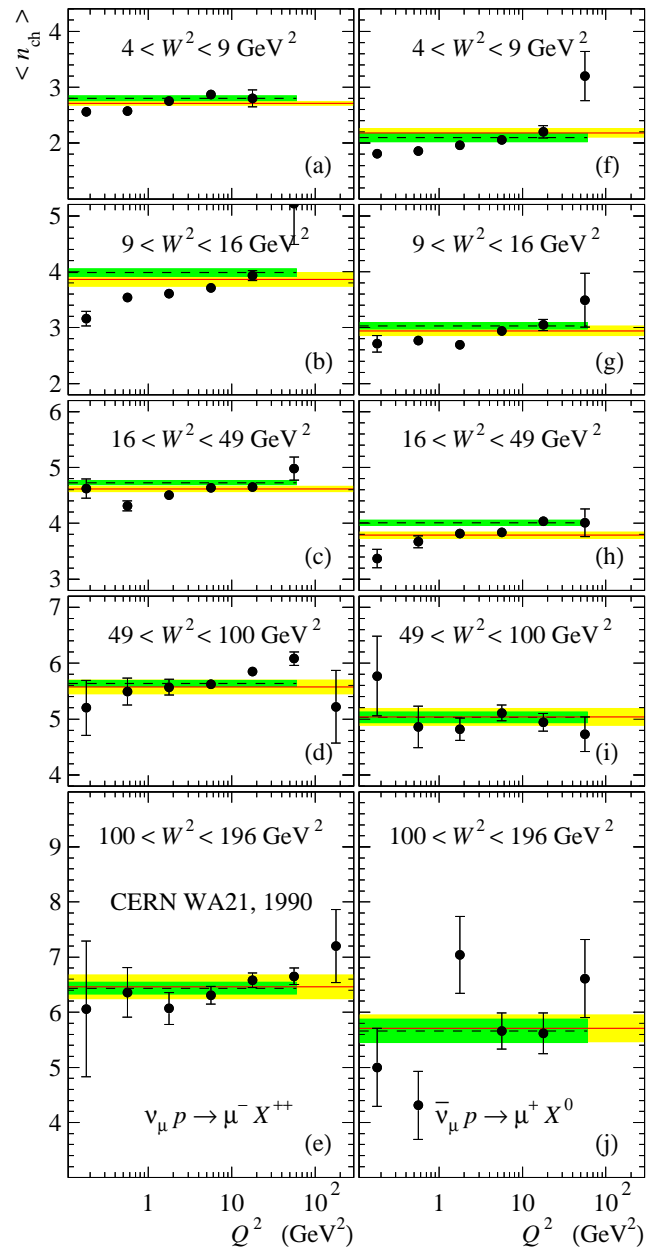}
 \caption{(Color online)
          Mean multiplicities of charged hadrons in $\nu_{\mu}p$ (left panels) and
          $\overline{\nu}_{\mu}p$ (right panels) CC interactions vs.\ $Q^2$ for various $W$
          slices measured in the experiment WA21 with the hydrogen filled BEBC bubble chamber
          at CERN \protect\cite{Jones:90} (filled circles). The errors shown are statistical only.
          For both $\nu_{\mu}p$ and $\overline{\nu}_{\mu}p$ reactions, the slices $4-5$~GeV
          and $5-7$~GeV used in Ref.~\protect\cite{Jones:90} are merged into the single one
          as is shown in panels (c) and (h). 
          The data of Ref.~\protect\cite{Jones:90} obtained at $W>14$~GeV are not shown
          because they are strongly affected by additional cutoff conditions~\protect\cite{Footnote-Jones:90}.
          The solid lines represent the overall averages for each $W$ slice.
          The dashed lines are the averages over the five $Q^2$ slices presented in
          Ref.~\cite{Jones:90} employing the same data sample and the same binning in $W$.
          The filled bands display the 68\% confidence intervals around the estimated averages.
         }
 \label{Fig:MCHMCC_Q2_CERN_WA21}
 \end{figure}

Figure~\ref{Fig:MCHMCC_Q2_CERN_WA21} shows the mean charged multiplicities for the ${\nu}p$
and $\overline{\nu}p$ reactions presented in Ref.~\cite{Jones:90} (CERN BEBC WA21)
as a function of $Q^2$ for several $W$ slices.
Our averaging over these slices is also shown in the figure, along with the averages
obtained over the five $Q^2$ slices $0-1$, $1-5$, $5-10$, $10-25$, and $25-60$ GeV$^2$,
presented in Ref.~\cite{Jones:90}. For this comparison we merged the $W$ bins $4-5$ and
$5-7$~GeV into the single bin $4-7$~GeV, as is shown in panels (c) and (h) of
Fig.~\ref{Fig:MCHMCC_Q2_CERN_WA21} \cite{Footnote-Jones:90}.
It is seen that, except for the bin $W=4-7$~GeV for the $\overline{\nu}p$ reaction,
the two methods of averaging are in reasonable agreement with each other, even regardless
the fact that the averaging over the $Q^2$ slices does not include the unavailable
contributions at $Q^2>60$~GeV$^2$.
For $W>3$~GeV, the averaged multiplicities shown in Fig.~\ref{Fig:MCHMCC_Q2_CERN_WA21}
are also in good agreement with the corresponding measurements from Ref.~\cite{Jones:92},
which do not include the data from the lowest bin $W=2-3$~GeV. The latter is however
important for our aims and must be incorporated into the data set for fitting.

The earlier results of Ref.~\cite{Allen:81} (CERN BEBC WA21, ${\nu}p$) are based on about
two-thirds of the data published in Ref.~\cite{Grassler:83} and the latter is in turn superseded
by the final statistics of the WA21 experiment used in Refs.~\cite{Jones:90,Jones:92} after
reprocessing with another method for treatment of systematic effects. The improved method led to
considerably lower values of the charged hadron multiplicities at high $W$ and consequently to lower
slopes for both ${\nu}p$ and $\overline{\nu}p$ reactions (see Table~\ref{Tab:MCHMCC_collection}).
However the ${\nu}p$ data from Ref.~\cite{Allen:81} obtained in the resonance region
($W\lesssim 2$~GeV) were not incorporated into the three latter analyses presented in
Refs.~\cite{Grassler:83,Jones:90,Jones:92}.
Considering that the results of all four analyses of the WA21 data sample in the overlapping
mid-$W$ region agree with each other within the statistical errors, one might take it that
the data of Ref.~\cite{Allen:81} at $W<2$~GeV are not stale and hence we can safely
add this low-$W$ subsample into the set for fitting, along with the full data sample
at higher $W$.
To sum up, in the subsequent analysis we utilize the CERN~WA21 data at $W<2$~GeV, $W=2-3$~GeV,
and $W>3$~GeV from, respectively, Refs.~\cite{Allen:81}, \cite{Jones:90}, and \cite{Jones:92}.

With the arguments similar to those used for the WA21 experiment, we include into the data set for fitting the low-$W$ (resonance region)
data subsample of Ref.~\cite{Barlag:82} (CERN BEBC WA25, $\overline{\nu}n$) which is not overruled
by the final statistics result of the WA25 experiment reported in Ref.~\cite{Jongejans:89}.
We note that the earlier WA25 data from Ref.~\cite{Allasia:80} obtained with lower statistics
are in quite good agreement with those from Refs.~\cite{Barlag:82,Jongejans:89} within the
overlapping region $W\gtrsim2$~GeV.

\clearpage 

\section{The fitting procedure}
\label{Sec:TheFittingProcedure}

As is known, the mean charged-hadron multiplicity in hadron-hadron, lepton-hadron
and $e^+e^-$ collisions grows faster than logarithmically with $W$ or $\sqrt{s}$
(the c.m.\ energy); at $\sqrt{s}>30-40$~GeV $\langle{n_{\text{ch}}\rangle}$ can be
well parametrized by the expressions $a+b\ln s+c\ln^2s$, $a+bs^n$, $a+b\exp(c\sqrt{\ln\,s})$,
etc., thus confirming the perturbative QCD predictions.

The invariant hadronic mass values available in the $\nu/\overline{\nu}$ experiments
discussed in Sect.~\ref{Sec:Experimental_data} are essentially lower than that in the
experiments with the $p/\overline{p}$, ${\mu}^\pm$ and $e^\pm$ beams.
It is stated in some papers (see, e.g., Refs.~\cite{Kittel:04,GrosseOetringhaus:10}
and references therein) that the energy dependence of the charged-hadron multiplicity
is almost universal irrespective of the nature of the projectile.
Our statistical analysis only partially confirms this assertion which is based
on by-eye comparison of conflicting and sometimes obsolete data. 

The analysis also shows that the parametrizations of the above kinds (with obvious
substitution $s\to W^2$) can not provide satisfactory fits to the neutrino data
in the whole $W$ range. 
The simplest expression \eqref{StandardParametrization} with target dependent
slope and intercept matches well the neutrino data at sufficiently large $W$
as well as the antineutrino data at any $W$.
On the other hand, neither logarithmic nor power-low parametrizations
describe the neutrino data at small and intermediate $W$.
An appropriate but yet simple expression is the following combination
of two polynomials in $\ln{W}^2$: 
\begin{equation}
\label{n_c_gen}
\langle{n_{\text{ch}}}\rangle = \left\{
\begin{aligned}
a_1+b_1\ln X+c_1\ln^2 X & \enskip \text{for} \enskip X \le X_0, \\
a_2+b_2\ln X+c_2\ln^2 X & \enskip \text{for} \enskip X  >  X_0.
\end{aligned}
\right.
\end{equation}
Here $X=W^2/W_1^2$, $X_0=W_0^2/W_1^2$, the parameter $W_1$ is
the minimal allowed value of $W$
($W_1=m_p+m_\pi$, $m_n$, $m_p$, and $m_n+m_\pi$ for, respectively,
${\nu}p$, $\overline{\nu}p$, ${\nu}n$, and $\overline{\nu}n$).
The values of the parameters $a_i$, $b_i$, $c_i$ ($i=1,2$), and $W_0$ are subject
to determination by a statistical data analysis, conditions of smooth joining of
the branches \eqref{n_c_gen} in the point $W=W_0$ defining the transition boundary
between the resonance and deep-inelastic (DIS) regions, and certain additional
constraints (explained later).
Except for the special test fits (see below), we assume that
$a_1=\langle{n_{\text{ch}}}\rangle_{\min}=|Q_h|$, where $|Q_h|=2, 0, 1$, and $1$
for, respectively, the ${\nu}p$, $\overline{\nu}p$, ${\nu}n$, and $\overline{\nu}n$
reactions.
To avoid violation of the rule $\langle{n_{\text{ch}}}\rangle \ge |Q_h|$, we apply
the conditional minimum chi-square estimation under the restriction $b_1\ge0$.
The assumed continuity of $\langle{n_{\text{ch}}}\rangle$ and  
$\partial\langle{n_{\text{ch}}}\rangle/\partial\ln X$ in the point $X=X_0$
provides the relations
\begin{gather*}
a_1+ b_1\ln X_0+c_1\ln^2 X_0 = a_2+ b_2\ln X_0+c_2\ln^2 X_0, \\
b_1+2c_1\ln X_0              = b_2+2c_2\ln X_0,
\end{gather*}
which allow us to exclude any two parameters from the fit.
The particular choice of these dependent parameters is a matter of convenience.
We provisionally retain the term $c_2\ln^2X$ at $X>X_0$ to make certain that
(in accord with the conventional parametrization) for all reactions the
coefficient $c_2$ is compatible with zero within the statistical accuracy.

In the statistical analysis of the data given below we use the CERN function minimization
and error analysis package ``MINUIT'' (version 94.1) \cite{MINUIT}, taking care of getting
an accurate correlation matrix.
Unless otherwise indicated, the quoted errors of the output parameter correspond
to the usual one-standard-deviation ($1\sigma$) errors (MINUIT's default).
Reducing of the number of the independent free parameters simplifies
determination of their best-fit values and errors, but somewhat complicates
estimation of the errors in the slave parameters. 
The total uncertainties $\delta_{\pm}\langle{n_{\text{ch}}}\rangle$ of
$\langle{n_{\text{ch}}}\rangle=\langle{n_{\text{ch}}}\rangle_{\boldsymbol{\xi}}$
are determined through variation of the parameters  $\{a_{i}, b_{i},\ldots\}=\boldsymbol{\xi}$
around the best-fit values $\{\overline{a}_{i}, \overline{b}_{i},\ldots\}=\boldsymbol{\overline{\xi}}$
within the estimated bounds of uncertainty,
\begin{gather*}
\delta_+\langle{n_{\text{ch}}}\rangle =
  \max\,\langle{n_{\text{ch}}}\rangle_{\boldsymbol{\xi}}
       -\langle{n_{\text{ch}}}\rangle_{\overline{\boldsymbol{\xi}}},  \\ 
\delta_-\langle{n_{\text{ch}}}\rangle =
        \langle{n_{\text{ch}}}\rangle_{\overline{\boldsymbol{\xi}}}
 -\min\,\langle{n_{\text{ch}}}\rangle_{\boldsymbol{\xi}},
\end{gather*}
with the slave parameters varied within the corresponding $1\sigma$ or $2\sigma$ marginalized
confidence contours.

Typical systematic uncertainties in the experiments described in Sect.~\ref{Sec:Experimental_data}
are smaller than or comparable to the statistical ones. To achieve a conservative estimation
of the errors in the required parametrization of $\langle{n_{\text{ch}}\rangle}$, in all cases
where the authors do not provide the systematic errors of the measurements, we set these to
be equal to the statistical errors, as is common practice.

As the first step, we applied the most general ansatz \eqref{n_c_gen} for fitting to the
experimental data for each projectile and target. 
Consequently, we obtained that
\begin{itemize}
\item[  (i)] the branch at $X>X_0$ is not needed for the reactions with antineutrinos (hence $a_2=b_2=c_2=0$ 
             and the parameter $W_0$ is irrelevant for this case);
\item[ (ii)] for the ${\nu}p$ reactions (both for hydrogen and deuterium targets) the parameter $c_2$
             is compatible with zero within at least three standard deviations, and hence we set $c_2=0$ below;
\item[(iii)] the parameter $b_1$ is fully compatible with zero for ${\nu}p$ (both for hydrogen and deuterium targets)
             and ${\nu}n$ reactions, with a typical error of about $10^{-3}$ or less, hence we neglect $b_1$
             for these reactions.
\end{itemize}

In Sect.~\ref{Sec:Comparison} below we perform a detailed comparison of our fits with the experimental
data and also with relevant outputs of several modern Monte Carlo simulations.
Namely, we consider available predictions of
GENIE (Generates Events for Neutrino Interaction Experiments) \cite{Andreopoulos:2009rq}, 
NuWro (Wroc{\l}aw neutrino event generator) \cite{Sobczyk:2008zz},
GiBUU (Giessen Boltzmann-Uehling-Uhlenbeck) transport model \cite{Lalakulich:2013tca},
and of the multistring MC code VENUS (Very Energetic NUclear Scattering) \cite{Werner:1993uh}.
All these models are currently in use for the data processing and analysis of the neutrino oscillation experiments.
Before we proceed further, a brief mention should be made of the physics content of the models.
Only those features are discussed herein which are directly concerned with the present study.
Detailed descriptions of the GENIE, NuWro, and GiBUU hadronization models, can be found in
Refs.~\cite{Yang:09,Andreopoulos:2006-2009,Nowak:2006sx}, and \cite{Buss:2011mx}, respectively.

\section{Neutrino MC generators}
\label{Sec:Generators}

\subsection{GENIE}

In the resonance region, $W<1.7$~GeV, GENIE \cite{Andreopoulos:2006-2009,Andreopoulos:2009rq}
uses the simplified Rein-Sehgal model \cite{Rein:81} with 16 baryon resonances whose contributions
are added incoherently, along with a small fraction of the DIS contribution.
Above $1.7$~GeV, the generator uses the Andreopoulos-Gallagher-Kehayias-Yang (AGKY)
KNO hadronization model \cite{Yang:09} based on the DIS contribution, by integrating
an empirical low-$W$ model with PYTHIA\,6.4/JETSET routines \cite{Sjostrand:2006za}
at higher $W$.
The non-resonance multi-hadron production is modeled in a few steps.
As the very first step, the code computes the average charged-hadron multiplicity
using the expression~\eqref{StandardParametrization} with the coefficients determined
from the FNAL E545 \cite{Zieminska:83} for ${\nu}p$ and ${\nu}n$ interactions and
CERN~WA25 \cite{Barlag:82} for $\overline{\nu}p$ and $\overline{\nu}n$ interactions
(recall that both experiments used the deuterium filled bubble chambers).
The average hadron multiplicity is then computed as $1.5\langle{n_{\text{ch}}}\rangle$,
according to the BEBC WA59 data \cite{Wittek:1988ke} on ${\nu}\text{Ne}$
and $\overline{\nu}\text{Ne}$ CC interactions.
At the next step, the actual hadron multiplicity is generated assuming that the
multiplicity dispersion is described by the Koba-Nielsen-Olesen (KNO) scaling
relation \cite{Koba:1972ng},
$
{\langle{n}\rangle}P_n(s)=\psi\left(n/{\langle{n}\rangle}\right),
$
where $P_n(s)$ is the probability of generating $n$ hadrons and $\psi(z)$ is
a $s$-independent universal function parametrized as
$
\psi(z)=2e^{-c}c^{cz+1}/\Gamma(cz+1),
$
with the input parameter $c$ determined from the KNO-scaling distributions measured
in the same deuterium experiments \cite{Zieminska:83} and \cite{Barlag:82} for,
respectively, neutrino and antineutrino interactions.

\subsection{NuWro}

The NuWro generator \cite{Nowak:2006xv,Sobczyk:2008zz} shares many common features
with NEUT, GENIE, NUANCE, and FLUKA, but uses its own hadronization model.
For description of the low-$W$ region (below $1.6$~GeV in the current version), only
the $\Delta$ resonance is treated explicitly (with several options for the electromagnetic
form factors and with the axial form factor obtained as a fit to ANL and BNL data).
The heavier resonances are assumed to enter as an average background of the DIS contribution,
via quark-hadron duality \cite{NuWro-QHD}.
The DIS structure functions are described using Bodek-Yang low-$Q^2$ corrections \cite{Bodek:2002vp}.
The PYTHIA\,6.1 fragmentation routines \cite{Sjostrand:2000wi} are used for the quark-level
simulation of the final state formation at the invariant hadronic masses down to the single-pion
production threshold. The KNO scaling relation is not used.
Five input parameters of PYTHIA\,6.1 are adjusted for better agreement with the measured charged
multiplicities from Refs.~\cite{Zieminska:83,Barlag:82} (deuterium) and \cite{Grassler:83} (hydrogen).

\subsection{GiBUU}

The GiBUU transport model \cite{Lalakulich:2013tca,Buss:2011mx,Leitner:2008ue} is a sophisticated
multipurpose theoretical tool which includes the neutrino-induced reactions as an option.
The model is based on coupled semi-classical kinetic equations describing
the space-time evolution of many-particle systems under the influence of mean-field potentials
and collision terms. In the case of neutrino-nucleon/nucleus collisions, the initial state of
a hadronic system is obtained via external models: at energies below a few hundred of GeV,
the hadrons propagate in mean fields and scatter according to cross sections; at higher energies,
the concept of pre-hadronic interactions is implemented to account for color transparency and
formation-time effects.
The GiBUU code is able to incorporate all possible resonances provided that the form factors are available,
but to the moment it includes contributions from 13 resonances with the invariant masses below 2~GeV.
The vector form factors of these resonances are taken from the recent Mainz-Dubna Unitary Isobar Model
(MAID\,2005) analysis for the helicity amplitudes (see references in Refs.~\cite{Buss:2011mx,Leitner:2008ue}).
The axial couplings are obtained from the PCAC relation. The axial form factor for $\Delta$
is refitted to the ANL data.
The non-resonant pion background is modeled phenomenologically by a technique based on
invariant amplitudes taken from MAID, as described in Ref.~\cite{Leitner:2008ue}.
The non-vector background contributions are fitted to the ANL data for the total pion production
cross sections.
No data on hadron multiplicities are used as input, hence these can be used for validation of the model. 

\subsection{VENUS}

The VENUS model \cite{Werner:1993uh} is based on Gribov-Regge theory of multiple Pomeron exchange and
classical relativistic string dynamics, and is closely related to the dual parton model and quark-parton
string model. While the VENUS model is primarily designed to treat nuclear collisions at ultrarelativistic
energies, it includes the neutrino-nucleon interactions as a by-product option which only uses the fragmentation
facilities of the VENUS code, and is only applicable at sufficiently high $s$ and $W$.

VENUS does not explicitly utilize the KNO-scaling hypothesis but calculations of the multiplicity distributions
with VENUS show the KNO scaling in agreement with the data.
A large amount of $e^+e^-$ and lepton-nucleon data were used to adjust parameters and validate the model
but, to our knowledge, the data on the hadron multiplicities in the ${\nu}N/\overline{\nu}N$ collisions
were not used in this adjustment.

\section{Comparison with data}
\label{Sec:Comparison}

\begin{figure*}[htb]
 \includegraphics[width=0.97\linewidth]{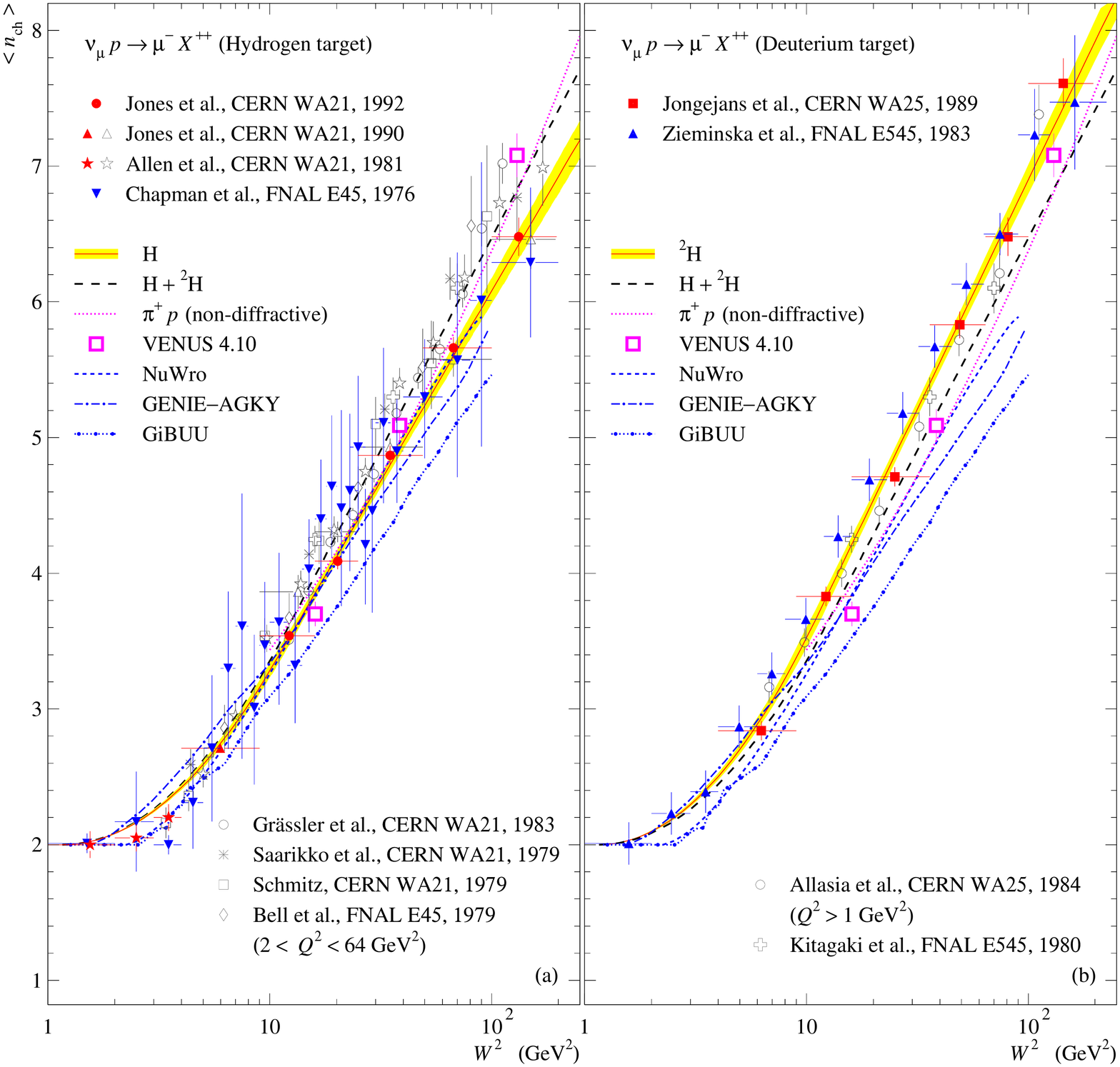}
 \caption{(Color online)
          A comparison between the fitted and measured charged-hadron multiplicity vs.\ $W^2$
          for the reaction $\nu_{\mu}p\to\mu^{-}X^{++}$ in hydrogen (a) and deuterium (b).
          The data points are from the experiments
          FNAL~E45 \protect\cite{Chapman:76,Bell:79},
          FNAL~E545 \protect\cite{Kitagaki:80,Zieminska:83},
          CERN~WA21 \protect\cite{Saarikko:79,Allen:81,Grassler:83,Jones:90,Jones:92}, and
          CERN~WA25 \protect\cite{Allasia:84,Jongejans:89}.
          The vertical error bars represent the quadratic sum of the statistical and systematic errors.
          Only the points marked by filled symbols are included into the analysis;
          the others are shown for comparison. 
          Solid curves enclosed by the $1\sigma$ confidence bands are calculated with
          the parameters from fits (A) (see Table~\protect\ref{Tab:MCHMCC_mn_p}).
          The long-dashed curves show the formal fit (A) to the combined set of the hydrogen
          and deuterium data (``H+$^2$H'' column in Table~\protect\ref{Tab:MCHMCC_mn_p}).
          The large open squares show the VENUS\,4.10 model prediction~\protect\cite{Werner:1993uh}.
          The curves marked ``GENIE/AGKY'', ``NuWro'', and ``GiBUU'' are borrowed from 
          Refs. \protect\cite{Andreopoulos:2009rq}, \protect\cite{Sobczyk:2008zz},
          and \protect\cite{Lalakulich:2013tca}, respectively.
          Calculations in Refs.~\protect\cite{Werner:1993uh,Andreopoulos:2009rq,Sobczyk:2008zz,Lalakulich:2013tca}
          are done for a free proton target and are shown in panel (b) only for comparison.
          The dotted curves in both panels represent a fit to the charged-hadron multiplicity for
          the non-diffractive component of the $\pi^+p$ reactions \protect\cite{Bardadin-Otwinowska:82}.
          }
 \label{Fig:MCHMCC_W2_mn_p}
 \end{figure*}

\begin{figure*}[htb]
 \includegraphics[width=0.97\linewidth]{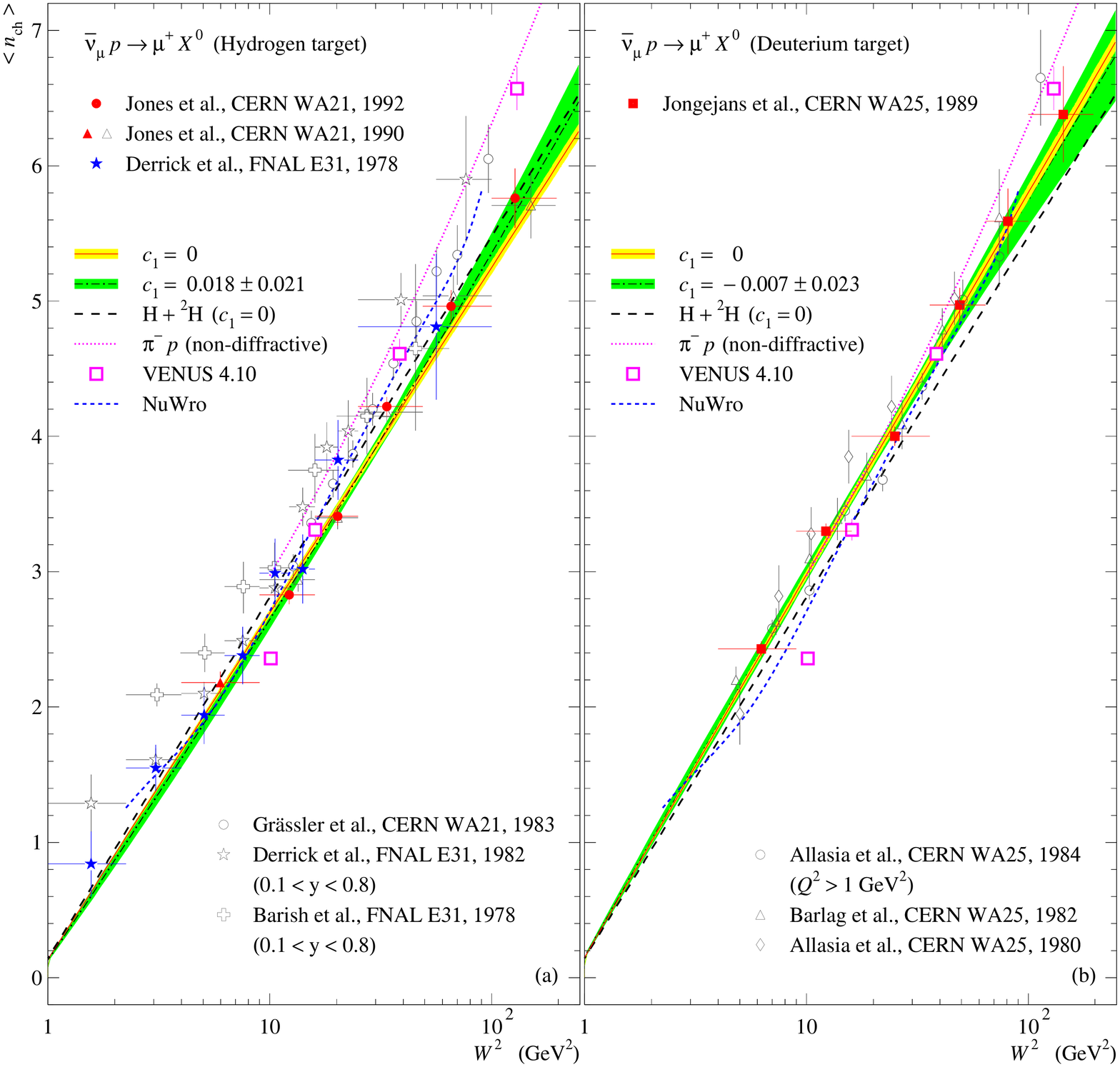}
 \caption{(Color online)
          A comparison between the fitted and measured charged-hadron multiplicity vs.\ $W^2$
          for the reaction $\overline{\nu}_{\mu}p\to\mu^{+}X^{0}$ in hydrogen (a) and deuterium (b).
          The data points are from the experiments
          FNAL~E31 \protect\cite{Derrick:78,Barish:78,Derrick:82},
          CERN~WA21 \protect\cite{Grassler:83,Jones:90,Jones:92}, and
          CERN~WA25 \protect\cite{Allasia:80,Barlag:82,Jongejans:89}.
          The vertical error bars represent the quadratic sum of the statistical and systematic errors.
          Only the points marked by filled symbols are included into the analysis;
          the others are shown for comparison. 
          Solid and dashed-dotted curves enclosed by the $1\sigma$ confidence bands are calculated with
          the parameters from fits (A) and (B), respectively (see Table~\protect\ref{Tab:MCHMCC_ma_p}).
          The long-dashed curves show the formal fit (A) to the combined set of the hydrogen
          and deuterium data (``H+$^2$H'' column in Table~\protect\ref{Tab:MCHMCC_ma_p}).
          The large open squares show the VENUS\,4.10 model prediction~\protect\cite{Werner:1993uh}.
          The dashed curves marked ``NuWro'' are borrowed from Ref.~\protect\cite{Sobczyk:2008zz}.
          Both VENUS and NuWro calculations are performed for a free proton target and are shown
          in panel (b) only for comparison.
          The dotted curves in both panels represent a fit to the charged-hadron multiplicity for
          the non-diffractive component of the $\pi^-p$ reactions \protect\cite{Bardadin-Otwinowska:82}.
          }
 \label{Fig:MCHMCC_W2_ma_p}
 \end{figure*}

\begin{figure*}[htb]
 \includegraphics[width=0.97\linewidth]{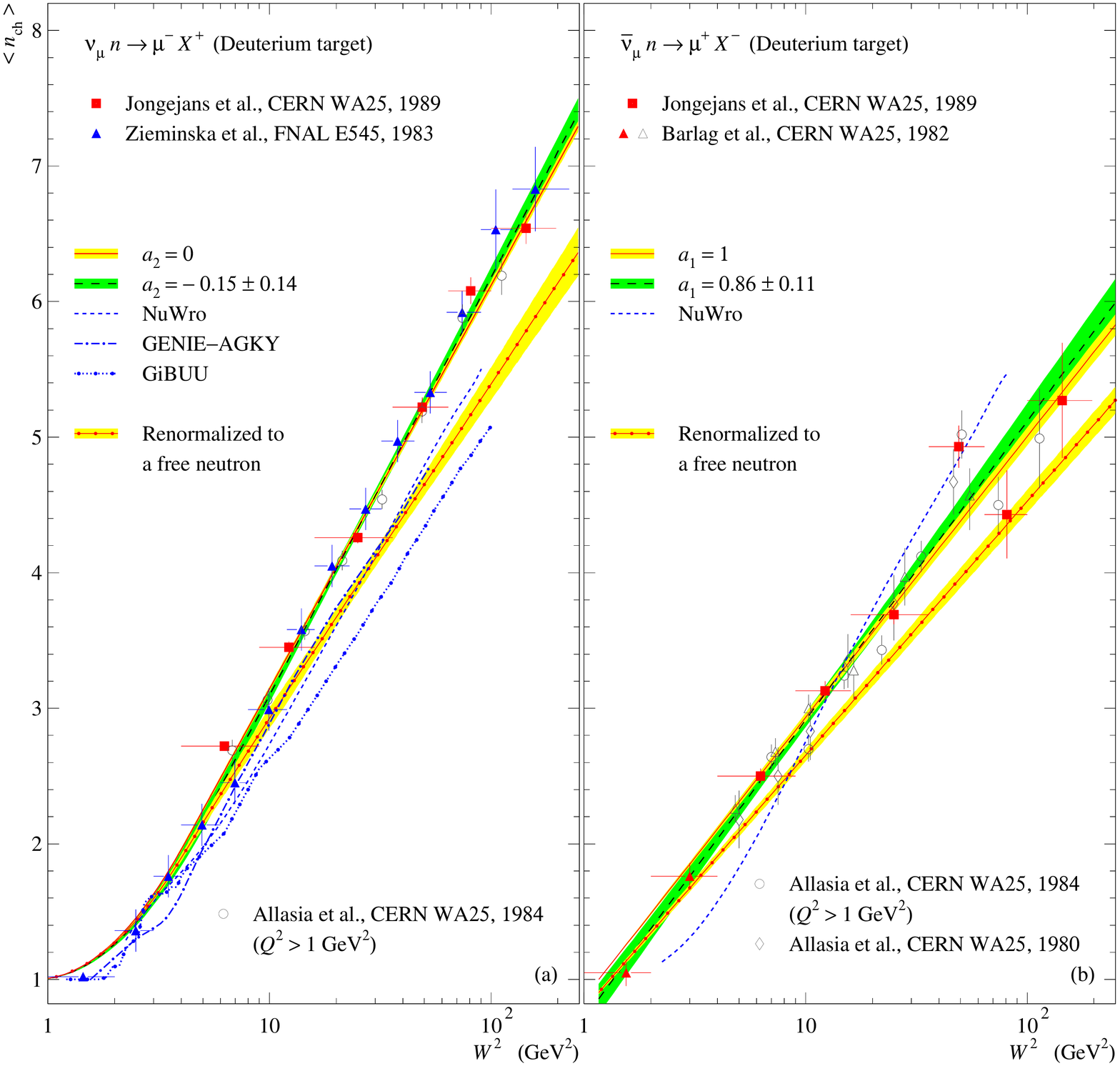}
 \caption{(Color online)
          A comparison between the fitted and measured charged-hadron multiplicities vs.\ $W^2$
          for the reactions $\nu_{\mu}n\to\mu^{-}X^{+}$ (a) and $\overline{\nu}_{\mu}n\to\mu^{+}X^{-}$ (b).
          The data points are from the experiments
          FNAL~E545 \protect\cite{Zieminska:83} and
          CERN~WA25 \protect\cite{Allasia:80,Barlag:82,Allasia:84,Jongejans:89}.
          The vertical error bars represent the quadratic sum of the statistical and systematic errors.
          Only the points marked by filled symbols are included into the analysis;
          the others are shown for comparison. 
          The solid and long-dashed curves enclosed by the $1\sigma$ confidence bands
          are calculated with the parameters of the fits (A) and (B), respectively
          (see Tables \protect\ref{Tab:MCHMCC_mn_n} and \protect\ref{Tab:MCHMCC_ma_n}).
          Also shown are the curves and the corresponding $1\sigma$ confidence bands obtained according
          to Eq.~\protect\eqref{bound-free} by using the parameters of the default fits (see the main text). 
          The curves marked ``GENIE/AGKY'', ``NuWro'', and ``GiBUU'' are borrowed from 
          Refs.~\protect\cite{Andreopoulos:2009rq}, \protect\cite{Sobczyk:2008zz},
          and \protect\cite{Lalakulich:2013tca}, respectively, where the calculations are performed for
          a free neutron target.
         }
 \label{Fig:MCHMCC_W2_mn_n_ma_n}
 \end{figure*}

The individual fits were performed in several versions for each reaction and
the main results of these fits are presented in Tables \ref{Tab:MCHMCC_mn_p}--\ref{Tab:MCHMCC_ma_n}
and in Figs. \ref{Fig:MCHMCC_W2_mn_p}--\ref{Fig:MCHMCC_W2_mn_n_ma_n}.
The fitted parameters in the tables are shown with a certain excess of accuracy
in order to avoid discontinuity in the joining of the branches \eqref{n_c_gen}.
Note, besides, that at least three digits in the mantissas of these parameters are needed
for an accurate representation of the confidence bands and error contours displayed in
Figs.~\ref{Fig:MCHMCC_W2_errors_mn_p}--\ref{Fig:MCHMCC_W2_errors_ma_n}.

The general notation used in Figs. \ref{Fig:MCHMCC_W2_mn_p}--\ref{Fig:MCHMCC_W2_mn_n_ma_n} is as follows:
the filled symbols  denote the data involved into the statistical analysis,
while the open symbols are for the data which do not satisfy the selection criteria discussed
in length in Sect.~\ref{Sec:Experimental_data} (these data are shown for completeness and comparison purposes).
The vertical error bars include both statistical and systematic uncertainties added in quadrature.
The horizontal bars display the $W^2$ bins; they are shown only for the data points involved into
the present analysis. The other nomenclature is explained in the legends and captions of the figures.
Below, in this section, we discuss in more detail the results of our analysis for each reaction type.

\subsection{\protect ${\nu}p$}
\label{Sect:nu-p}

The best-fit parameters for $\langle{n^{{\nu}p}_{\text{ch}}}\rangle$ are listed in Table~\ref{Tab:MCHMCC_mn_p}
for the two cases, when the parameter $a_1$ is set to 2 (A) or remains unfixed (B).
\begin{table}[htb]
\caption{Best-fit parameters for the ${\nu}p$ reaction, obtained from the H, $^2$H, and combined
         H+$^2$H data sets.
         In fit (A) the value of $a_1$ is set to 2, while in fit (B) it remains a free parameter.
         In both fits, $b_1=c_2=0$ and $W_1=m_p+m_\pi$. Parameter $W_0$ is in GeV.
         }
\label{Tab:MCHMCC_mn_p}
\center{
\begin{tabularx}{\linewidth}{c>{\centering}Xccc}                                                               \hline\hline                    \noalign{\smallskip}
\#                   & Param.\              & H dataset          & $^2$H dataset      & H+$^2$H dataset     \\ \noalign{\smallskip}\hline      \noalign{\smallskip}
\multirow{5}*{(A)}   & $c_1$                & $0.277\pm0.011$    & $0.329\pm0.015$    & $0.292\pm0.008$     \\ \noalign{\smallskip}
                     & $a_2$                & $0.665\pm0.157$    & $0.362\pm0.207$    & $0.421\pm0.133$     \\ \noalign{\smallskip}
                     & $b_2$                & $1.215\pm0.053$    & $1.468\pm0.065$    & $1.358\pm0.043$     \\ \noalign{\smallskip}
                     & $W_0^2$              & $10.46\pm1.76$     & $10.82\pm2.02$     & $11.90\pm1.48$      \\ \noalign{\smallskip}
                     & \chiNDF              & \F{39.5}{32}{1.23} & \F{23.1}{18}{1.28} & \F{207.6}{52}{3.99} \\ \noalign{\smallskip}\hline      \noalign{\smallskip}
\multirow{6}*{(B)}   & $a_1$                & $1.862\pm0.082$    & $1.980\pm0.167$    & $1.893\pm0.075$     \\ \noalign{\smallskip}
                     & $c_1$                & $0.311\pm0.031$    & $0.334\pm0.046$    & $0.315\pm0.024$     \\ \noalign{\smallskip}
                     & $a_2$                & $0.678\pm0.194$    & $0.372\pm0.267$    & $0.452\pm0.171$     \\ \noalign{\smallskip}
                     & $b_2$                & $1.212\pm0.065$    & $1.465\pm0.083$    & $1.345\pm0.053$     \\ \noalign{\smallskip}
                     & $W_0^2$              & $8.181\pm2.185$    & $10.43\pm3.99$     & $9.910\pm2.216$     \\ \noalign{\smallskip}
                     & \chiNDF              & \F{29.0}{31}{0.94} & \F{23.1}{17}{1.36} & \F{200.4}{51}{3.93} \\ \noalign{\smallskip}\hline\hline\noalign{\smallskip}
\end{tabularx}
}
\end{table}
The fits are performed separately for the hydrogen (H) and deuterium ($^2$H) data sets as well as for the
combined H+$^2$H data set.
Both (A) and (B) fits produce satisfactory correlation matrices and comparable values of the parameters
$c_1$, $a_2$, $b_2$, and $W_0$ for the given target type. Regardless of the fact that the resulting
$\chi^2/\text{NDF}$ value is somewhat better in case (B), the latter is less preferred since it violates
the rule $\langle{n_{\text{ch}}}\rangle\ge|Q_h|$ for the hydrogen data at about the $2\sigma$ level.
Note that exclusion of the data points of Ref.~\cite{Chapman:76} from the set for fitting has
little impact on the parameter values shown in the second column of Table~\ref{Tab:MCHMCC_mn_p},
but would slightly increase the errors of the parameters and corresponding $\chi^2/$NDF.
The same remains true after a re-sampling of the data of Ref.~\cite{Chapman:76}.
The (A) and (B) fits to the combined set of the hydrogen and deuterium data
(last column in Table~\ref{Tab:MCHMCC_mn_p}) yield unacceptably large values of $\chi^2/\text{NDF}$
indicating a lack of coincidence between the H and $^2$H subsets.

\begin{figure}[htb]
 \includegraphics[width=0.97\linewidth]{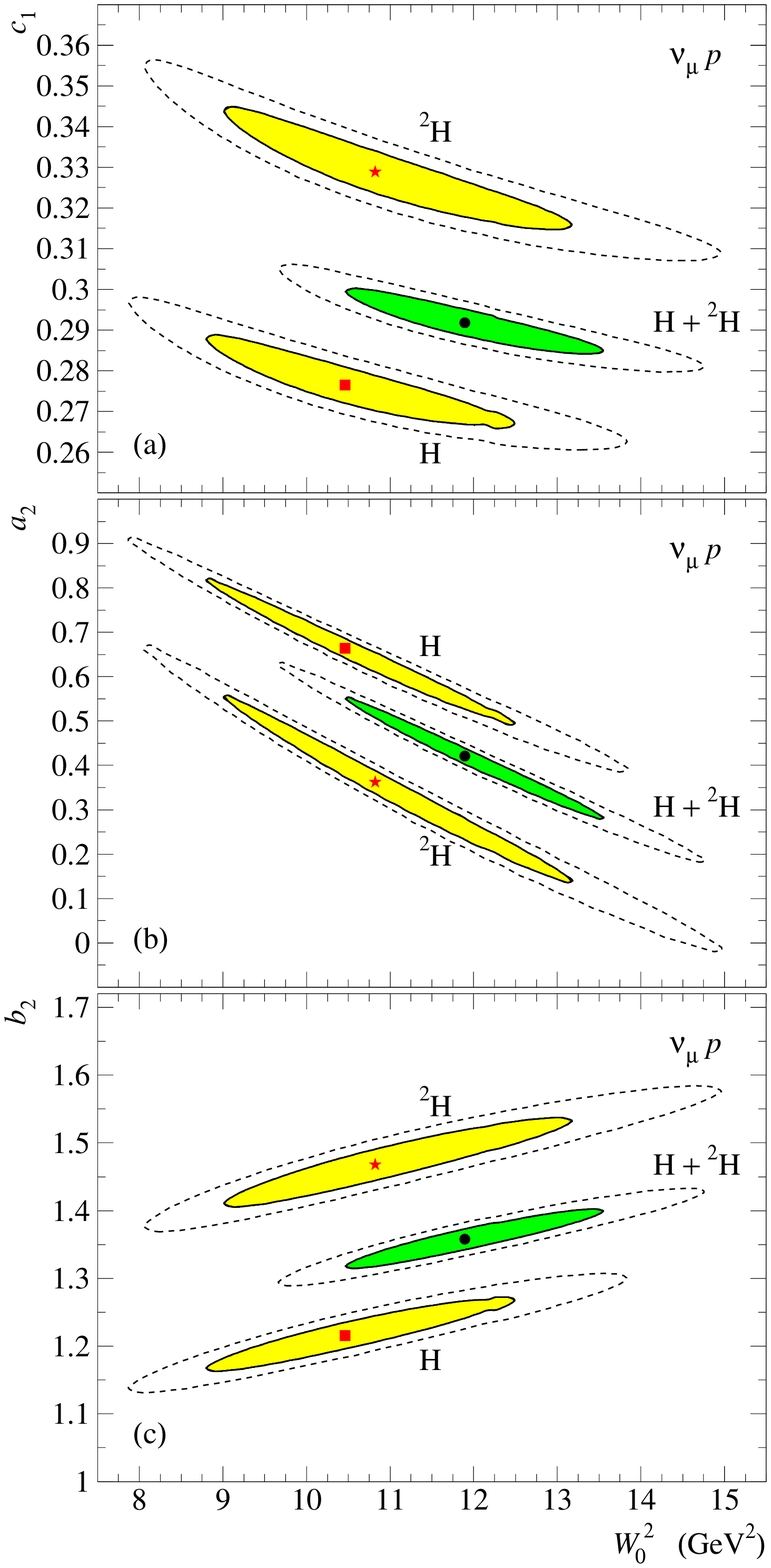}
 \caption{(Color online)
          Error contours for the three pairs of interdependent parameters
          listed in Table~\protect\ref{Tab:MCHMCC_mn_p}, version (A) for the
          ${\nu}p$ reaction and derived from the H, $^2$H, and H+$^2$H data sets.
          The solid and dashed contours indicate the $68$\% and $95$\%
          confidence levels, respectively.
          The points indicate the best-fit values of the parameters.
         }
 \label{Fig:MCHMCC_W2_errors_mn_p}
 \end{figure}

Figure \ref{Fig:MCHMCC_W2_errors_mn_p} shows the 68\% and 95\%~C.L.\ contours for the three
parameter pairs $(c_1,W_0^2)$, $(a_2,W_0^2)$, and $(b_2,W_0^2)$ evaluated for case (A).
It is seen that the best-fit parameters for the H and $^2$H targets are certainly incompatible and thus
the formal fit to the H+$^2$H data is meaningless, except for the region $W^2\lesssim10~\text{GeV}^2$.
In other words, the ${\nu}p$ charged multiplicities are undoubtedly different for
the hydrogen and deuterium targets.
The most natural explanation of this difference is the considerable effect of rescattering
inside the deuteron, widely discussed in the literature (see, e.g., Ref.~\cite{Rescattering}
and references therein).

A comparison between the data and fit (A) is shown in Fig.~\ref{Fig:MCHMCC_W2_mn_p}.
The $1\sigma$ uncertainty band around the solid curve is calculated by using the
confidence contours from Fig.~\ref{Fig:MCHMCC_W2_errors_mn_p}.
The figure also contains the $W$ dependencies of $\langle{n^{{\nu}p}_{\text{ch}}}\rangle$
predicted by the multistring MC code VENUS \cite{Werner:1993uh},
by the neutrino MC generators GENIE \cite{Andreopoulos:2009rq} and NuWro \cite{Sobczyk:2008zz},
and by the GiBUU transport model \cite{Lalakulich:2013tca}.
Calculations in Refs.~\cite{Werner:1993uh,Andreopoulos:2009rq,Sobczyk:2008zz,Lalakulich:2013tca}
were performed for a free proton target and are plotted in panel (b) only to emphasize their
non-applicability to the bound proton in deuterium.

Let us  note that in the current version of the GENIE generator \cite{Andreopoulos:2009rq},
the unphysical bulge in $\langle{n^{{\nu}p}_{\text{ch}}}\rangle$ is removed, which has occurred
within the intermediate $W$ range in earlier versions of the code, and presumably originated from
combining together the PYTHIA and KNO based hadronization models \cite{Sobczyk:2008zz}.
The NuWro code does not use the KNO scaling assumption and its predictions are smooth.
Both the GENIE and NuWro curves agree within the errors with the hydrogen data shown in
Fig.~\ref{Fig:MCHMCC_W2_mn_p}~(a). Besides, the NuWro prediction is in quite good agreement
with our best-fit band. Recall, however that the model has been fine-tuned so as to
bring its predictions closer to the measured charged multiplicities.

As is seen from Fig.~\ref{Fig:MCHMCC_W2_mn_p}~(a), the VENUS\,4.10 model predicts the steepest slope
which matches the earlier CERN~WA21 data \cite{Grassler:83} at high $W$, rather than the more recent
WA21 data of Ref.~\cite{Jones:92} (which primarily determine the shape of the fit).
The GiBUU model yields essentially lower charged multiplicity and a more flat slope at high $W$.
It is worth noting, however, that the corresponding $1\sigma$ confidence interval estimated
in Ref.~\cite{Lalakulich:2013tca} within GiBUU is very wide, especially at high $W$, and fully
covers both the best-fit band and data points. We also note that the GiBUU predictions \cite{Lalakulich:2013tca}
for the averaged multiplicities for neutral meson production on hydrogen and neon targets are
in better agreement with the data.

The VENUS, GENIE, NuWro, and GiBUU curves are systematically lower than the deuterium data points
shown in Fig.~\ref{Fig:MCHMCC_W2_mn_p}~(b), further indicating the essential difference between
the ${\nu}p$ charged-hadron multiplicities extracted from the hydrogen and deuterium experiments.
The dotted curves in Fig.~\ref{Fig:MCHMCC_W2_mn_p} show the fit to the energy
dependence of the average charged multiplicity for the non-diffractive component
of the $\pi^+p$ reactions,
$$
\langle{n^{\pi^+\!p}_{\text{ch}}}\rangle\strut_{\text{ND}}=1.98+0.31\ln s+0.14\ln^2s,
$$
obtained in Ref.~\cite{Bardadin-Otwinowska:82} (uncertainty of the fit is not provided by the authors).
The similarity of $\langle{n^{{\nu}p}_{\text{ch}}}\rangle$ and $\langle{n^{\pi^+\!p}_{\text{ch}}}\rangle_{\text{ND}}$
is expected from a simple quark-model consideration that $W^+p$ and $\pi^+p$ collisions must generate the same
quark-diquark string ($u-uu$) with charge 2.

\subsection{\protect $\overline{\nu}p$}
\label{Sect:nubar-p}

\begin{figure}[htb]
\includegraphics[width=0.97\linewidth]{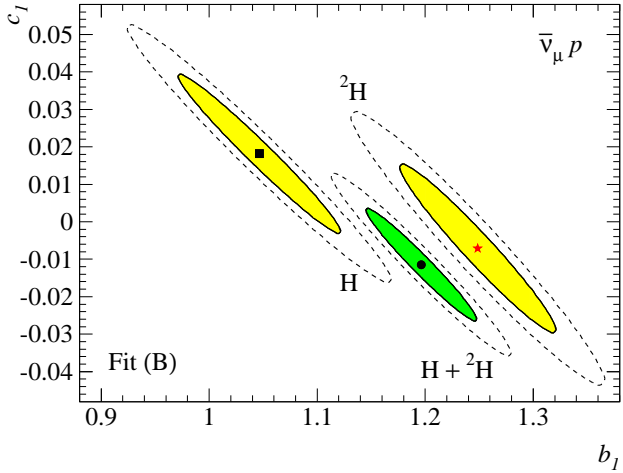}
\caption{(Color online)
         Error contours for the independent parameters $b_1$ and $c_1$
         listed in Table~\protect\ref{Tab:MCHMCC_ma_p}, version (B) for the
         $\overline\nu_{\mu}p$ reaction and derived from the H, $^2$H, and H+$^2$H
         data sets. The solid and dashed contours indicate the 68\% and 95\%
         confidence levels, respectively.
         The filled symbols indicate the best-fit values of the parameters
         in fit (B).
        }
\label{Fig:MCHMCC_W2_errors_ma_p_B}  
\end{figure}

\begin{figure}[htb]
\includegraphics[width=0.97\linewidth]{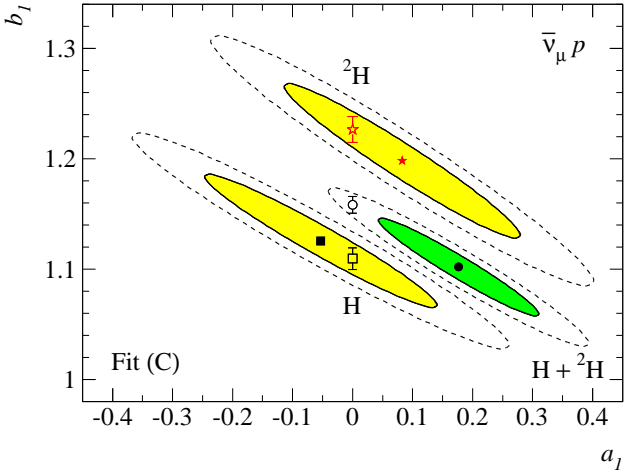}
\caption{(Color online)
         The same as in Fig.~ \ref{Fig:MCHMCC_W2_errors_ma_p_B} but for the
         pair of the parameters $a_1$ and $b_1$ listed in Table~\protect\ref{Tab:MCHMCC_ma_p},
         version (C). 
         The filled symbols indicate the best-fit values of the parameters
         in fit (C), while the open symbols show the best-fit values of $b_1$
         when the value of $a_1$ is set to zero according to fit (A).
        }
\label{Fig:MCHMCC_W2_errors_ma_p_C}  
\end{figure}

The best-fit parameters for $\langle{n^{\overline{\nu}p}_{\text{ch}}\rangle}$ are listed
in Table~\ref{Tab:MCHMCC_ma_p} in four versions of the fit: with one (A), two (B,C), and three (D)
uncorrelated free parameters $a_1$, $b_1$, and $c_1$; we recall that only one branch in
Eq.~\eqref{n_c_gen} is sufficient here.
\begin{table}[htb]
\caption{Best-fit parameters for the $\overline{\nu}p$ reaction, obtained from the H, $^2$H, and 
         combined H+$^2$H data sets.
         It is set
         $a_1=0$ in fits (A) and (B), and
         $c_1=0$ in fit (C). In all four fits, $a_2=b_2=c_2=0$ and $W_1=m_n$. Parameter $W_0$ is in GeV.
         }
\label{Tab:MCHMCC_ma_p}
\center{
\begin{tabularx}{\linewidth}{c>{\centering}Xccc}                                                             \hline\hline                    \noalign{\smallskip}
\#                    & Param.\             & H dataset          & $^2$H dataset     & H+$^2$H dataset    \\ \noalign{\smallskip}\hline      \noalign{\smallskip}
\multirow{2}*{(A)}    & $b_1$               & $1.110\pm0.010$    & $1.227\pm0.012$   & $1.158\pm0.008$    \\ \noalign{\smallskip}
                      & \chiNDF             & \F{9.98}{13}{0.77} & \F{4.38}{5}{0.88} & \F{71.9}{19}{3.79} \\ \noalign{\smallskip}\hline      \noalign{\smallskip}
\multirow{3}*{(B)}    & $b_1$               & $1.047\pm0.075$    &  $1.249\pm0.072$  & $ 1.196\pm0.051$   \\ \noalign{\smallskip}
                      & $c_1$               & $0.018\pm0.021$    & $-0.007\pm0.023$  & $-0.012\pm0.015$   \\ \noalign{\smallskip}
                      & \chiNDF             & \F{8.31}{12}{0.69} & \F{4.15}{4}{1.04} & \F{70.6}{18}{3.92} \\ \noalign{\smallskip}\hline      \noalign{\smallskip}
\multirow{3}*{(C)}    & $a_1$               & $-0.053\pm0.194$   & $0.083\pm 0.197$  & $0.177\pm0.134$    \\ \noalign{\smallskip}
                      & $b_1$               & $ 1.126\pm0.060$   & $1.198\pm 0.070$  & $1.102\pm0.044$    \\ \noalign{\smallskip}
                      & \chiNDF             & \F{9.81}{12}{0.82} & \F{3.97}{4}{0.99} & \F{68.0}{18}{3.78} \\ \noalign{\smallskip}\hline      \noalign{\smallskip}
\multirow{4}*{(D)}    & $a_1$               & $0.346\pm0.472$    & $0.287\pm0.861$   & $0.500\pm0.407$    \\ \noalign{\smallskip}
                      & $b_1$               & $0.829\pm0.312$    & $1.052\pm0.598$   & $0.866\pm0.276$    \\ \noalign{\smallskip}
                      & $c_1$               & $0.051\pm0.052$    & $0.024\pm0.099$   & $0.040\pm0.046$    \\ \noalign{\smallskip}
                      & \chiNDF             & \F{6.43}{11}{0.58} & \F{3.75}{3}{1.25} & \F{65.3}{17}{3.84} \\ \noalign{\smallskip}\hline\hline\noalign{\smallskip}
\end{tabularx}}
\end{table}
All these fits are repeated for the H, $^2$H, and H+$^2$H data sets.
Figures \ref{Fig:MCHMCC_W2_errors_ma_p_B} and \ref{Fig:MCHMCC_W2_errors_ma_p_C} display
the 68\% and 95\%~C.L.\ error contours for the two independent pairs of parameters,
$(b_1,c_1)$ and $(a_1,b_1)$ evaluated for the cases (B) and (C), respectively.

From Table~\ref{Tab:MCHMCC_ma_p} and Figs. \ref{Fig:MCHMCC_W2_errors_ma_p_B} and \ref{Fig:MCHMCC_W2_errors_ma_p_C}
it can be concluded the following:
\begin{itemize}
\item[  (i)] Similar parameters obtained in the different fits roughly
             (within about $2\sigma$) coincide for each data set.
\item[ (ii)] The coefficients $a_1$ and $c_1$ are well compatible with zero, namely,
             $|\overline{a}_1| \lesssim 0.7\sigma$, $|\overline{c}_1|\lesssim\sigma$
             for the H data set and
             $|\overline{a}_1| \lesssim 0.4\sigma$, $|\overline{c}_1|\lesssim 0.2\sigma$
             for the $^2$H data set.
             The values of $\chi^2/\text{NDF}$ only slowly vary with increasing
             the number of free parameters, while the errors in determination
             of the parameters quickly grow.
             Therefore even the simplest one-parameter fit (A) is quite appropriate
             for description of the available hydrogen and deuterium data.
\item[(iii)] The values of $\chi^2/\text{NDF}$ for the H+$^2$H data set are unacceptably
             large and the corresponding $2\sigma$ error contours do not intersect
             the H and $^2$H contours, suggesting that the $\overline{\nu}p$ charged
             multiplicities are strongly different for the hydrogen and deuterium targets
             -- exactly as in the case of ${\nu}p$ reaction and certainly by the same token.
\end{itemize}


Figure~\ref{Fig:MCHMCC_W2_ma_p} shows a comparison between the data and the fits (A) and (B).
The $1\sigma$ uncertainty bands around the corresponding curves are calculated by using the
confidence contours from Fig.~\ref{Fig:MCHMCC_W2_errors_ma_p_B}.
The data of FNAL~E31 from Refs.~\cite{Barish:78} and \cite{Derrick:78} are shown to demonstrate the effect
of the stringent cut-off conditions.
It is also shown the $W$ dependencies of $\langle{n^{\overline{\nu}p}_{\text{ch}}}\rangle$
predicted by the VENUS\,4.10 model \cite{Werner:1993uh} and by the NuWro neutrino generator \cite{Sobczyk:2008zz}
for a free proton target. At high $W$, both VENUS and NuWro match the earlier WA21 data of
Ref.~\cite{Grassler:83} but disagree with the updated WA21 data set \cite{Jones:92}.
The dotted curves in Fig.~\ref{Fig:MCHMCC_W2_ma_p} show the fit to the energy
dependence of the average charged multiplicity for the non-diffractive component
of the $\pi^-p$ reactions,
$$
\langle{n^{\pi^-\!p}_{\text{ch}}}\rangle\strut_{\text{ND}}=0.69+0.76\ln s+0.10\ln^2s,
$$
obtained in Ref.~\cite{Bardadin-Otwinowska:82} (uncertainty of the fit is not provided by the authors).
There is a sizable disagreement with respect to the best-fit curves for
$\langle{n^{\overline{\nu}p}_{\text{ch}}}\rangle$, which increases with energy.
We note thereupon that the statement of Ref.~\cite{Bardadin-Otwinowska:82} that, in accord with
the quark-model prediction (see, Sect.~\ref{Sect:nu-p}), the charged multiplicities in the non-diffractive
$\pi^+p/\pi^-p$ reactions agree well with these in the ${\nu}p/\overline{\nu}p$ reactions, was based,
in fact, on a comparison with the partially outdated data and/or with the data obtained under strong cut-off
conditions.

\subsection{\protect ${\nu}n$}
\label{Sect:nu-n}

\begin{figure}[htb]
 \includegraphics[width=0.97\linewidth]{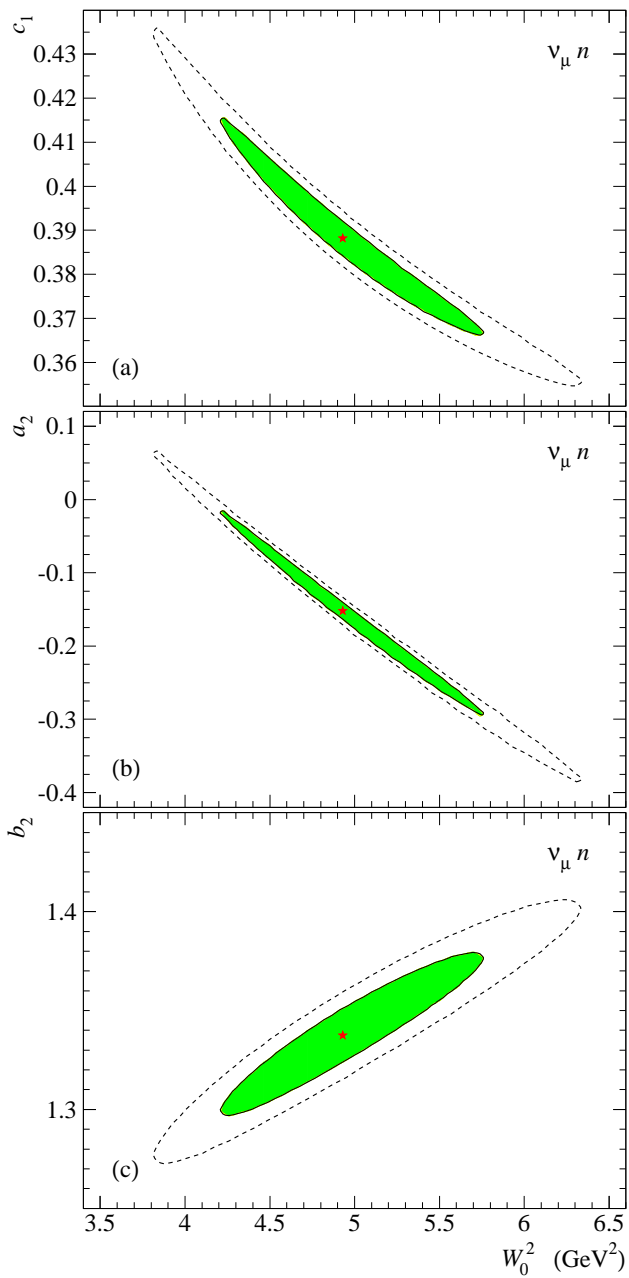}
 \caption{(Color online)
          Error contours for the three pairs of interdependent parameters
          listed in Table~\protect\ref{Tab:MCHMCC_mn_n}, version (B) for
          the $\nu_{\mu}n$ reaction.
          The solid and dashed contours indicate the 68\% and 95\%
          confidence levels, respectively.
          The points indicate the best-fit values of the parameters.
         }
 \label{Fig:MCHMCC_W2_errors_mn_n}
 \end{figure}

Table~\ref{Tab:MCHMCC_mn_n} shows the best-fit parameters for $\langle{n^{{\nu}n}_{\text{ch}}}\rangle$.
Here we consider two cases, when the parameter $a_2$ is set to zero (A) or remains unfixed (B).
In both these cases, we fix $a_1=1$ and $b_1=c_2=0$, since variations of these parameters
(separately or in any combination) result in worsening of the correlation matrix and/or in
a significant increase of the chi-square value.
\begin{table}[htb]
\caption{Best-fit parameters for the ${\nu}n$ reaction, obtained from the deuterium data set.
         In fit (B) the value of $a_2$ is set to 0, in fit (A) it remains a free parameter;
         in both fits $a_1=1$, $b_1=c_2=0$ and $W_1=m_p$. Parameter $W_0$ is in GeV.
         }
\label{Tab:MCHMCC_mn_n}
\center{
\begin{tabularx}{\linewidth}{c>{\centering}X>{\centering\arraybackslash}X}              \hline\hline\noalign{\smallskip}
    Parameter        &    (A)              &    (B)              \\ \noalign{\smallskip}\hline      \noalign{\smallskip}
$c_1$                & $0.418\pm0.004$     &  $0.388\pm0.024$    \\ \noalign{\smallskip}
$a_2$                & $0$                 & $-0.152\pm0.139$    \\ \noalign{\smallskip}
$b_2$                & $1.294\pm0.007$     &  $1.337\pm0.041$    \\ \noalign{\smallskip}
$W_0^2$              & $4.132\pm0.033$     & $4.930\pm0.766$     \\ \noalign{\smallskip}
\chiNDF              & \F{28.89}{18}{1.61} & \F{26.02}{17}{1.53} \\ \noalign{\smallskip}\hline\hline\noalign{\smallskip}
\end{tabularx}}
\end{table}
The regions of correlated errors for the parameter pairs $(c_1,W_0^2)$, $(a_2,W_0^2)$, $(b_2,W_0^2)$
for the fit (B) are shown in Fig.~\ref{Fig:MCHMCC_W2_errors_mn_n}, and comparison of the fits
(A) and (B) with the data from Refs.~\cite{Zieminska:83,Allasia:80,Barlag:82,Allasia:84,Jongejans:89}
is presented in Fig.~\ref{Fig:MCHMCC_W2_mn_n_ma_n}~(a).
As is seen from Table~\ref{Tab:MCHMCC_mn_n}, the values of $\chi^2/\text{NDF}$ are the same for the
(A) and (B) versions of the fit and are quite acceptable, considering the apparent inconsistency
of the E545 \cite{Zieminska:83} and W25 \cite{Jongejans:89} data for $W^2=4-10~\text{GeV}^2$
(see Fig.~\ref{Fig:MCHMCC_W2_errors_mn_n}~(a)) which negatively affects the goodness of the fits \cite{Footnote-WA25-E545}.
Although both fits in this region are not quite reliable, the obtained parameters are formally consistent
with each other within the errors. The errors in case (B) are very large, and besides, this fit cannot be safely
extrapolated to higher $W$ because of the negative value of $a_2$. So the fit (A) seems to be more preferable.

In order to confront our best fit for $\langle{n^{{\nu}n}_{\text{ch}}}\rangle$ with the results of
Refs.~\cite{Andreopoulos:2009rq}, \cite{Sobczyk:2008zz}, and \cite{Lalakulich:2013tca} obtained for
the free neutron, it is reasonable to imply that the relative nuclear corrections to the charged-hadron multiplicity
are, in the first approximation, the same for the neutrino scattering on proton and neutron for the same nuclear
target, namely
\begin{equation}
\label{bound-free}
\frac{\langle{n^{{\nu}p}_{\text{ch}}}\rangle\strut_{\text{free}}}{\langle{n^{{\nu}p}_{\text{ch}}}\rangle\strut_{\text{bound}}}
=
\frac{\langle{n^{{\nu}n}_{\text{ch}}}\rangle\strut_{\text{free}}}{\langle{n^{{\nu}n}_{\text{ch}}}\rangle\strut_{\text{bound}}}.
\end{equation}
The $W$ dependence of $\langle{n^{{\nu}n}_{\text{ch}}}\rangle_{\text{free}}$ evaluated in this approximation
is plotted in Fig.~\ref{Fig:MCHMCC_W2_errors_mn_n}~(a) together with the $1\sigma$ confidence interval; the charged multiplicities
$\langle{n^{{\nu}p}_{\text{ch}}}\rangle\mathstrut_{\text{free}} =\langle{n^{{\nu}p}_{\text{ch}}}\rangle\mathstrut_{\text{H}}$,
$\langle{n^{{\nu}p}_{\text{ch}}}\rangle\mathstrut_{\text{bound}}=\langle{n^{{\nu}p}_{\text{ch}}}\rangle\mathstrut_{\text{$^2$H}}$,
and
$\langle{n^{{\nu}n}_{\text{ch}}}\rangle\mathstrut_{\text{bound}}=\langle{n^{{\nu}n}_{\text{ch}}}\rangle\mathstrut_{\text{$^2$H}}$,
are evaluated using our default fits (A). 
It is seen that the GENIE, NuWro, and GiBUU predictions are in much better agreement with the ``renormalized''
best-fit multiplicity than with that for a bound neutron. 

\subsection{\protect $\overline{\nu}n$}
\label{Sect:nubar-n}

\begin{figure}[htb]
 \includegraphics[width=0.971\linewidth]{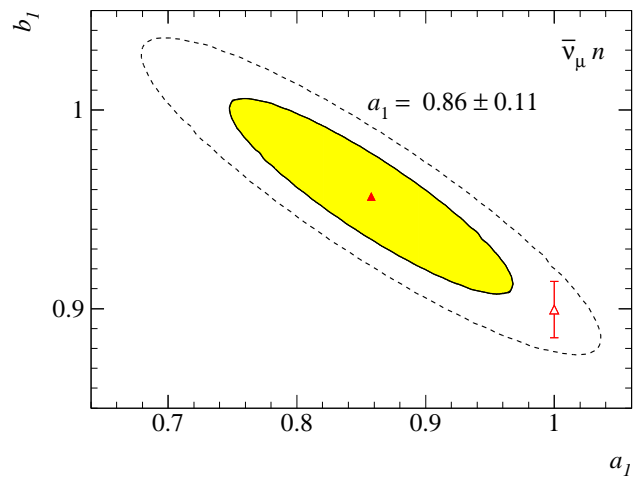}
 \caption{(Color online)
          Error contours for the independent parameters $a_1$ and $b_1$
          listed in Table~\protect\ref{Tab:MCHMCC_ma_n}, version (B) for the
          $\nu_{\mu}n$ reaction. 
          The solid and dashed contours indicate the 68\% and 95\%
          confidence levels, respectively.
          The filled triangle indicates the best-fit values of the parameters.
          The open triangle is for the best-fit value of $b_1$ when
          $a_1$ is set to 1 according to fit (A).
         }
 \label{Fig:MCHMCC_W2_errors_ma_n} 
 \end{figure}

Table~\ref{Tab:MCHMCC_ma_n} shows the best-fit parameters for $\langle{n^{\overline{\nu}n}_{\text{ch}}}\rangle$.
\begin{table}[htb]
\caption{Best-fit parameters for the $\overline{\nu}n$ reaction, obtained from the deuterium data set.
         In fit (A) the value of $a_1$ is set to 1, in fit (A) it remains a free parameter;
         in both fits $c_1=c_2=a_2=b_2=0$ and $W_1=m_n+m_\pi$. Parameter $W_0$ is in GeV.
         }
\label{Tab:MCHMCC_ma_n}
\center{
\begin{tabularx}{\linewidth}{c>{\centering}X>{\centering\arraybackslash}X}           \hline\hline\noalign{\smallskip}
   Parameter         &    (A)             &    (B)             \\ \noalign{\smallskip}\hline     \noalign{\smallskip}
$a_1$                & $1$                & $0.858\pm0.110$    \\ \noalign{\smallskip}
$b_1$                & $0.900\pm0.014$    & $0.957\pm0.049$    \\ \noalign{\smallskip}
\chiNDF              & \F{19.09}{7}{2.72} & \F{15.28}{6}{2.55} \\ \noalign{\smallskip}\hline\hline\noalign{\smallskip}
\end{tabularx}}
\end{table}
We again consider two cases, when the parameter $a_1$ is set to 1 (A) or remains unfixed (B).
In both cases, we fix $c_1=c_2=a_2=b_2=0$ since, due to the low amount of data points and their
scatter at high $W$, it is unreasonable to increase the number of free parameters.
Owing to the same reasons, the values of $\chi^2/\text{NDF}$ are large and it is difficult to make
a choice between the fits (A) and (B).
The 68\% and 95\%~C.L.\ error contours for the pair $(a_1,b_1)$ in the fit (B) are shown in
Fig.~\ref{Fig:MCHMCC_W2_errors_ma_n}.
It is seen that the parameters obtained in the fits (A) and (B) are compatible only within
the $2\sigma$ error ellipse.

A comparison of the fits (A) and (B) with the data from Refs.~\cite{Allasia:80,Barlag:82,Allasia:84,Jongejans:89}
is presented in Fig.~\ref{Fig:MCHMCC_W2_mn_n_ma_n}~(b).
Also shown are the NuWro generator prediction \cite{Sobczyk:2008zz} and the best fit recalculated with
the default parameters (A) by the same procedure as in the case of the ${\nu}n$ reaction (see Sect.\ \ref{Sect:nu-n}).
Unexpectedly, the disagreement of the NuWro curve with the renormalized best-fit $1\sigma$ confidence band
is even worse than that with the bound neutron and exceeds the fitting uncertainty caused by the data spread
in the high-$W$ region.
Notice that all the data shown in Fig.~\ref{Fig:MCHMCC_W2_mn_n_ma_n}~(b) were obtained in different stages
of the same experiment and independent measurements are needed to resolve the discrepancies and improve
statistical significance of the fit.

\section{Additional test}

As a useful cross-check, in this section, we will try to extrapolate our results to the FNAL~E632
measurements of $\langle{n^{{\nu}N}_{\text{ch}}}\rangle$ \cite{Vataga:2000ux,Vataga:1997cn,Vataga:1997PhD}
performed with the 15 foot Bubble Chamber exposed to a wide-band beam of muon (anti)neutrinos from Tevatron.
The average energy for the $\nu_{\mu}$ interactions in this experiment was about 145~GeV,
providing $W$ up to about 25~GeV.
The kinematic cuts $Q^2>1~\text{GeV}^2$, $W>2~\text{GeV}$ leave 4476 $\nu\text{Ne}$ CC events
from 5567 of the total sample.
In the two data runs (1985 and 1987) the chamber was filled with a neon-hydrogen mixture containing
75\% and 63\% molar neon of density of 0.71~g/cm$^3$ and 0.54~g/cm$^3$, respectively.

\begin{figure}[htb]
 \includegraphics[width=0.96\linewidth]{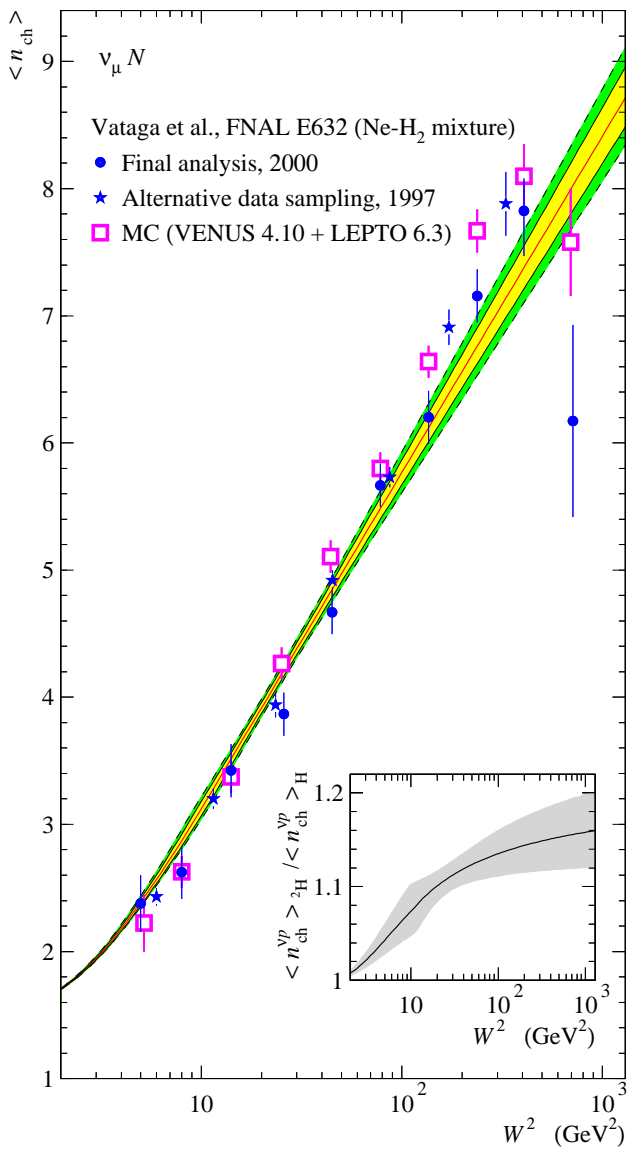}
 \caption{(Color online)
          Comparison of the FNAL~E632 data \protect\cite{Vataga:2000ux} (filled circles) with
          the parametrization \protect\eqref{Ne-H} (solid curve with bands).
          The thin solid and dashed curves indicate the 68\% and 95\%~C.L., respectively.
          Also shown are the results of an earlier analysis of the same data sample performed
          with a different sampling strategy \protect\cite{Vataga:1997cn} (filled stars) and
          of a Monte Carlo simulation performed with VENUS\,4.10 and LEPTO\,6.3 (open squares).
          The error bars on the experimental data points represent the statistical errors only. 
          The insert shows the ratio of the charged multiplicities $\langle{n^{{\nu}p}_{\text{ch}}}\rangle$
          on the deuterium and hydrogen targets calculated from our best fits (A) with
          the parameters listed in Table~\protect\ref{Tab:MCHMCC_mn_p};
          the gray band indicates the estimated uncertainty of the ratio (at 68\%~C.L.).
         }
 \label{Fig:MCHMCC_W2_mn_Ne-H_2_Vataga_FNAL_E632}
 \end{figure}

A neutrino-nucleus collision can be treated as a sequential process in which the neutrino interacts
with a single nucleon and then the resulting secondaries move through nuclear matter initiating
an intranuclear cascade.
A number of selection criteria were applied in the E632 data processing and analysis in order to separate
the ``pure'' (cascade-free) ${\nu}N$ interactions.
The measured $W$ dependence of $\langle{n^{{\nu}N}_{\text{ch}}}\rangle$ is shown in
Fig.~\ref{Fig:MCHMCC_W2_mn_Ne-H_2_Vataga_FNAL_E632}.

The VENUS\,4.10 \cite{Werner:1993uh} and LEPTO\,6.3 \cite{Ingelman:1992} codes were used
in the E632 analysis to simulate the processes of fragmentation and rescattering, and lepton-nucleon collisions.
Since the selection criteria are model dependent, the stability of the result was checked
by variations of the parameters in the VENUS model.

The best-fit parametrizations of the world data on the ${\nu}p$ and ${\nu}n$ charged-hadron
multiplicities obtained in the previous sections allow us to describe the multiplicities
of the cascade-free events measured in the experiment E632.
Let $\langle n_\text{ch}^{\nu(A,Z)}\rangle$ be the mean charged-hadron multiplicity for
the reaction ${\nu}_{\mu}+(A,Z)\to{\mu}X$, where $(A,Z)$ is the nucleus containing $Z$ protons
and $A-Z$ neutrons. Since the intranuclear cascade events are already subtracted from the E632
data sample, we can consider the nucleus as a superposition of free nucleons. Then
\begin{equation}
\label{n(A,Z)_def}
\langle n_\text{ch}^{\nu(A,Z)}\rangle
= \frac{\sum_kk\left[Z\sigma_k({\nu}p)+(A-Z)\sigma_k({\nu}n)\right]}
{Z\sigma_{\text{tot}}({\nu}p)+(A-Z)\sigma_{\text{tot}}({\nu}n)},
\end{equation}
where $\sigma_k({\nu}p)$ and $\sigma_k({\nu}n)$ are the cross sections for the production
of $k$ charged hadrons in the exclusive ${\nu}p$ and ${\nu}n$ reactions, respectively, while
$\sigma_{\text{tot}}({\nu}p)$ and $\sigma_{\text{tot}}({\nu}n)$ are the total cross sections.
Rewriting Eq.~\eqref{n(A,Z)_def} in terms of the elementary multiplicities $\langle{n^{{\nu}p}_{\text{ch}}}\rangle$
and $\langle{n^{{\nu}n}_{\text{ch}}}\rangle$ we get
\begin{equation}
\label{n(A,Z)}
\langle n_\text{ch}^{\nu(A,Z)}\rangle =
 \frac{\langle{n^{{\nu}p}_{\text{ch}}}\rangle}{1+\dfrac{(A-Z)r}{Z}}
+\frac{\langle{n^{{\nu}n}_{\text{ch}}}\rangle}{1+\dfrac{Z}{(A-Z)r}},
\end{equation}
where $r=\sigma_{\text{tot}}({\nu}n)/\sigma_{\text{tot}}({\nu}p)$. 
Considering now that
\begin{itemize}
 \item[  (i)] the total cross-section ratio $r$ very slowly evolves at high neutrino energies, remaining
              close to 2 (the value expected from the naive parton model) within at least a few percent
              for $E_{\nu}=15-300$~GeV (see, e.g., Ref.~\cite{Kuzmin:2006dt} 
              and references therein),
 
\item[ (ii)] $\langle{n^{{\nu}p}_{\text{ch}}}\rangle \approx \langle{n^{{\nu}n}_{\text{ch}}}\rangle$
             at the energies under consideration, and

\item[(iii)] the E632 neon-hydrogen target is almost isoscalar, 
\end{itemize}
we can simplify Eq.~\eqref{n(A,Z)} as follows:
\begin{equation}
\label{n_ch_free}
\langle{n^{\nu(A,Z)}_{\text{ch}}}\rangle \approx \langle\overline{n}_{\text{ch}}\rangle\left(1-\frac{\kappa}{6}\right)
\left[1-\frac{\kappa\left(r-2+4\delta\right)}{9}\right].
\end{equation}
Here
\begin{gather*}
\langle\overline{n}_{\text{ch}}\rangle = \frac{\langle{n^{{\nu}p}_{\text{ch}}}\rangle+\langle{n^{{\nu}n}_{\text{ch}}}\rangle}{2}, \\
\kappa = \frac{\langle{n^{{\nu}p}_{\text{ch}}}\rangle-\langle{n^{{\nu}n}_{\text{ch}}}\rangle}{\langle\overline{n}_{\text{ch}}\rangle}, \\
\delta = \frac{A}{2Z}-1,
\end{gather*}
and inessential higher-order terms are omitted.
Taking into account that $\kappa \ll 1$ and $\delta \ll 1$, it is clear from Eq.~\eqref{n_ch_free} that
the allowed variations of $r$ would only negligibly affect the mean multiplicity and hence it is safe to set $r=2$.
Adopting these simple considerations for the E632 target, we have to take into account the fractional abundances
of the neon isotopes (close to natural: 90.48\%, 0.27\%, and 9.45\% for, respectively, $^{20}$Ne, $^{21}$Ne, and $^{22}$Ne)
and relative number of the CC events in the two runs (0.579 and 0.421, according to Ref.~\cite{Vataga:1997PhD}).
Finally, by using Eq.~\eqref{n(A,Z)} or \eqref{n_ch_free} we obtain the following approximate expression for
the charged multiplicity in the cascade-free ${\nu}N$ interactions with the E632 neon-hydrogen target:
\begin{equation}
\label{Ne-H}
\langle{n^{{\nu}N}_{\text{ch}}}\rangle\strut_\text{Ne-H$_2$} \approx
0.53\langle{n^{{\nu}p}_{\text{ch}}}\rangle\strut_\text{free}+
0.47\langle{n^{{\nu}n}_{\text{ch}}}\rangle\strut_\text{free},
\end{equation}
where the charged multiplicity on the free neutron target, $\langle{n^{{\nu}n}_{\text{ch}}}\rangle\strut_\text{free}$,
has to be calculated according to Eq.~\eqref{bound-free} with the parameters listed in Table~\protect\ref{Tab:MCHMCC_mn_p}
for our default fit (A).
A comparison of our prediction with the data is shown in Fig.~\ref{Fig:MCHMCC_W2_mn_Ne-H_2_Vataga_FNAL_E632}.
Open squares in the figure show the result of the Monte Carlo calculations of Ref.~\cite{Vataga:2000ux}
based on the LEPTO\,6.3 and VENUS\,4.10 codes. As in the case of ${\nu}p$ and $\overline{\nu}p$ reactions
(see Sections \ref{Sect:nu-p} and \ref{Sect:nubar-p}),
the VENUS model slightly overestimates the charged multiplicity at $W^2\gtrsim20~\text{GeV}^2$ leading
to a steeper slope. Since the VENUS model was extensively used in the E632 data analysis, the resulting
data may contain systematic biases. Accordingly, we can only conclude that there is at least a qualitative
agreement between the measured trend and our prediction based on Eqs.~\eqref{Ne-H} and \eqref{bound-free}.

\section{Conclusions}
\label{Sec:Conclusions}

In this paper, we suggest simple parametrizations for the mean charged-hadron
multiplicities (as functions of the invariant mass of the final hadron system, $W$)
in the charged-current neutrino and antineutrino interactions with hydrogen
and deuterium targets.
The parametrizations work rather well for the whole kinematic range of $W$
from the reaction threshold to the deep-inelastic region, and can be recommended
for use as inputs and/or validation tool in the modern neutrino Monte Carlo generators.
The simplest versions (A) (based on the minimal number of the fitted parameters)
are as a rule preferable.

Our statistical analysis of available consistent data unambiguously demonstrates that
both $\nu_{\mu}p$ and $\overline{\nu}_{\mu}p$ charged multiplicities (as functions
of $W$) are essentially different for the hydrogen and deuterium targets and thus only
the hydrogen data (and corresponding parametrizations) can be used for description
of the charged multiplicities for a free proton target.
Presented comparison with the results from several neutrino MC generators
and the cross-check of our best-fit parametrizations with the highest $W$ data
from the FNAL~E632 experiment show that the simple relation \eqref{bound-free}
can be used for estimating the charged multiplicity for the free neutron target.

In the deep-inelastic region, all the multiplicities follow simple linear
in $\ln W$ dependencies with strongly different slopes related to the
given projectile and nuclear target.
The data provide no evidence for an increase of the slopes with $W$
observed in hadron-hadron, lepton-hadron, or $e^+e^-$ collisions.
However, the currently accessible energies in the (anti)neutrino experiments
are not high enough to make an unambiguous extrapolation above $W=20-25$~GeV.  
In order to further improve the accuracy of the fitted parameters,
new dedicated experiments are needed.

\begin{acknowledgments}

This work was supported by the Federal Target Program ``Scientific and 
Scientific-Pedagogical Personnel of the Innovative Russia'' under Contracts 
No.~2012-1.5-12-000-1011-008 and 14.U02.21.0913, and by the Russian Foundation
for Basic Research, under Grant No.~10-02-00395-a.
The authors would like to thank S.~R.~Mishra, O.~V.~Teryaev, and E.~S.~Vataga
for useful discussions.

\end{acknowledgments}


\begin{thebibliography}{99}
\bibitem{Footnote-Jones:90}
 We do not analyze here the $Q^2$ slices plotted in Figs.~3b and 4b of Ref.~\protect\cite{Jones:90}
 for $W>14$~GeV, since the rejected region $Q^2>60$~GeV$^2$ strongly affects the mean charged multiplicity
 at such high values of $W$.
\bibitem{Footnote-WA25-E545}
  In order to somewhat improve the goodness of the fits, we merged the two poorly-consistent
  data points from  WA25 and E545, namely, the WA25 point in the lowest $W$ bin and the underlying
  (5th) E545 point, see Fig.~\ref{Fig:MCHMCC_W2_errors_mn_n}~(a).
  The corresponding NDF is therefore decremented by 1.
\end{thebibliography}
\end{document}